\documentclass[aoas,preprint]{imsart}

\RequirePackage[OT1]{fontenc}
\RequirePackage{amsthm,amsmath}
\RequirePackage{amsmath, amssymb}
\RequirePackage{natbib}
\RequirePackage[colorlinks,citecolor=blue,urlcolor=blue]{hyperref}
\RequirePackage[dvipdfmx]{graphicx}
\RequirePackage{bm}
\RequirePackage{pdflscape}

\arxiv{arXiv:0000.0000}

\startlocaldefs
\numberwithin{equation}{section}
\theoremstyle{plain}

\endlocaldefs

\begin{document}

\begin{frontmatter}
\title{Space and circular time log Gaussian Cox processes with application to crime event data}
\runtitle{Space and circular time log Gaussian Cox processes}

\begin{aug}
\author{\fnms{Shinichiro} \snm{Shirota}\ead[label=e1]{ss571@stat.duke.edu}}
\and
\author{\fnms{Alan} \snm{E. Gelfand}\ead[label=e2]{alan@stat.duke.edu}}
\runauthor{S. Shirota and A. E. Gelfand}

\affiliation{Department of Statistical Science, Duke University}

\address{Department of Statistical Science \\
Duke University \\
Durham, NC, 27708-0251\\
\printead{e1}\\
\phantom{E-mail:\ }\printead*{e2}}
\end{aug}

\begin{abstract}
We view the locations and times of a collection of crime events as a space-time point pattern.  So, with either a nonhomogeneous Poisson process or with a more general Cox process, we need to specify a space-time intensity.  For the latter, we need a \emph{random} intensity which we model as a realization of a spatio-temporal log Gaussian process. Importantly, we view time as circular not linear, necessitating valid separable and nonseparable covariance functions over a bounded spatial region crossed with circular time.  In addition, crimes are classified by crime type.  Furthermore, each crime event is recorded by day of the year which we convert to day of the week marks.

The contribution here is to develop models to accommodate such data. Our specifications take the form of hierarchical models which we fit within a Bayesian framework.  In this regard, we consider model comparison between the nonhomogeneous Poisson process and the log Gaussian Cox process.  We also compare separable vs. nonseparable covariance specifications.

Our motivating dataset is a collection of crime events for the city of San Francisco during the year 2012.  We have location, hour, day of the year, and crime type for each event.  We investigate models to enhance our understanding of the set of incidences.
\end{abstract}


\end{frontmatter}

\section{Introduction}
The times of crime events can be viewed as circular data.  That is, working at the scale of a day, we can imagine event times as wrapped around a circle of circumference $24$ hours (which, without loss of generality, can be rescaled to $[0, 2\pi)$).  Furthermore, over a specified number of days, we can view the set of event times, consisting of a random number of crimes, as a point pattern on the circle.  Suppose, additionally, that we attach to each crime event its spatial location over a bounded domain.  Then, for a bounded spatial region, we have a space-time point pattern over this domain, again with time being circular.

The contribution here is to develop suitable models for such data, motivated by a set of crime events for the city of San Francisco in $2012$.  The challenges we address involve (i) clustering in time - event times are not uniformly distributed over the $24$ hour circle; (ii) spatial structure - evidently, some parts of the city have higher incidence of crime events than others; (iii) crime type - characterization of point pattern varies with type of crime so different models are needed for different crime types; (iv) incorporating covariate information - we anticipate that introducing suitable \emph{constructed} spatial and temporal covariates will help to explain the observed point patterns; (v) the need for spatio-temporal random effects - the constructed spatial and temporal covariates will not adequately explain the space-time point patterns; (vi) the availability of marks - in addition to a location and a time within the day, each event has an associated day of the year which we convert to a day of the week.  We propose a range of point pattern models to address these issues; fortunately, our motivating dataset is rich enough to investigate them.

We focus on the problem of building a log Gaussian Cox process (LGCP) which includes, as a special case, a nonhomogeneous Poisson process (NHPP), over space and circular time.  We need to build a suitable intensity surface which is driven by a realization of a log Gaussian process incorporating a valid covariance function over space and time.

An initial model for a set of points in space and circular time is a nonhomogeneous Poisson process (NHPP) with an intensity $\lambda(\bm{s},t)$ over say $D \times S^1$ where $D$ is the spatial region of interest and time lies on the unit circle, $S^1$.  We illuminate such intensities below but we also note that an NHPP will not prove rich enough for our data. So, we propose a space by circular time log Gaussian Cox process (LGCP).  This leads to consideration of space-time dependence.  Does time of day affect the spatial pattern of crime?  Does location affect the point pattern of event times?  Hence, we consider both separable and nonseparable models.  We develop a parametric nonseparable space by circular time correlation function building on Gneiting's specification (see, \cite{Gneiting(02)}).   We note very recent work from \cite{Porcuetal(15)} which presents valid covariance functions on $R^1$ crossed with spheres.

Typically, time is modeled linearly, leading to a large literature on point patterns over bounded time intervals (see, e.g., \cite{DaleyVereJones(03)} and \cite{DaleyVereJones(08)}). Adding space,  \cite{BrixDiggle(01)} offer development of a space-time LGCP. \cite{RodriguesDiggle(12)} consider a space-time process convolution model for modeling of space time crime events.
\cite{Liangetal(14)} consider process convolution for space with a dynamic model for time.
\cite{Taddy(10)} proposes a Bayesian semiparametric framework for modeling correlated time series of marked spatial Poisson processes with application to tracking intensity of violent crime.

In fact, in this context, it is important to articulate the difference between viewing time in a \emph{linear} manner vs. a \emph{circular} manner.  With linear time there is a past and a future.  We can condition on the past and predict the future, we can incorporate seasonality and trend in time.  With circular time, as with angular data in general, we only obtain a value once we supply an orientation, e.g., the customary midnight with time, although, below, we argue to start the day at 02:00.  So, we have no temporal ordering of our crime events except within a defined 24 hour window.  We are only interested in modeling the intensity over space and circular time.  For us, prediction would consider the number of events of a particular crime type, in a specified neighborhood, over a window of hours during the day, adjusting for day of the week.  For a decision maker, the value would be to facilitate making daily spatial staffing decisions during 24 hour cycles.  We do not assert that one modelling approach is better than the other.  Rather, the modeling approaches address different questions and yield different inference.  We do note that our approach is novel in considering space with circularity of time.

There is a useful literature modeling crime data as linear in time, using past locations to predict future locations.  In this regard, see \cite{Mohleretal(11)}, \cite{Mohler(13)} and \cite{Chaineyetal(08)}. \cite{Mohleretal(11)} employ self-exciting point process models, similar to those used in earthquake modeling (e.g., \cite{Ogata(98)}), arguing that crime, when viewed linearly in time, can exhibit ``contagion-like'' behavior. \cite{Mohler(13)} consider a self-exciting process with background rate driven by a logGaussian Cox process to disentangle contagion from other types of correlation.
\cite{Chaineyetal(08)} focus on hotspot assessment.  This is purely spatial analysis, which may be implemented across various time periods for comparison.

Wrapping time to a circle takes us to the realm of directional or angular data where we find applications to, for example, wind directions, wave directions, and animal movement directions.  For a review of directional data methodology see, e.g., \cite{Fisher(93)}, \cite{JammSen(01)}, \cite{Mardia(72)} and \cite{MardiaJupp(00)}.  Traditional models for directional data have employed the von Mises distribution but recent work has elaborated the virtues of the wrapping and projection approaches, particularly in space and space-time settings (see \cite*{JonaGelfandJona(12)} and \cite{WangGelfand(14)}).

For event times during a day, wrapping time seems natural.  Again, these times only arise given an orientation.  However, crimes at 23:55 and 00:05 are as temporally close as crimes at 23:45 and 23:55.  Another example analogous to our setting might be to model the arrival times (over 24 hours) of patients to a hospital (and, to add space, we might consider the addresses of the arrivals).

Our data consists of a set of crime events in San Francisco (SF) during the year $2012$.  Each event has a time of day and a location. In fact, we also have a classification into crime type and we also have assignment of each crime to a district, arising by suitable partitioning of the city.  Lastly, we know the day of the year for the event, enabling consideration of day of the week effects.

There is a substantial literature which employs regression models to explain the incidence of crime using a variety of socio-economic variables.  In particular, for spatially referenced covariates, we can imagine employing census unit risk factors such as percent of home ownership, median family income, measures of neighborhood quality, along with racial and ethnic composition.  Such covariates could be developed as tiled spatial surfaces over San Francisco, using census data at say tract or block scale.  As an alternative, we introduce illustrative point-referenced constructed covariates in space and time.  In particular, with regard to space, we imagine high risk locations, so-called \emph{crime attractors}, i.e., places of high population and high levels of human activity such as commercial centers and malls.  Then, we view risk as exposure in terms of distance from such landmarks.  We adopt these covariates in both the NHPP and LGCP models.  As a temporal risk factor, we introduce into the mean a function which reflects the fact that, depending upon the type of crime, evening and late night hours may experience higher incidence of crime than morning and afternoon hours. Again, we adopt this covariate in both the NHPP and LGCP models. Exploratory data analysis in Section 2 reveals a day of the week effect with regard to daily crime time.

Finally, we address model comparison.  In particular, within the Bayesian framework, how do we compare a NHPP model with a LGCP model?  We adopt the strategy proposed in \cite{LeiningerGelfand(16)}.  Briefly, the idea is to develop a cross-validation, employing a fitting/training point pattern and a testing/validation point pattern. Using the validation point pattern, with regard to model adequacy, we look at empirical coverage vs. nominal coverage of credible predictive intervals.  In particular, these intervals are associated with the posterior distribution of predictive residuals for cell counts for randomly selected sets. With regard to model comparison, we look at rank probability scores (see, e.g., \cite{GneitingRaftery(07)} and \cite{Czadoetal(09)}) for the posterior distributions associated with predictive residuals, again for cell counts for randomly selected sets.

The format of the paper is as follows.
In section 2, we provide the details of our crime event dataset along with some exploratory analysis. In section 3, we consider model construction with associated theoretical background. Employing day of the week marks, in section 4, we introduce our model specifications and fitting strategies. In section 5, we provide the inference results for both a simulated dataset and the crime event dataset. Finally, in Section 6, we present a brief summary along with proposed future work.

\section{The dataset}
\begin{figure}[htbp]
  \caption{The map of San Francisco (left) and crime counts on each day of week (right)}
 \begin{minipage}{0.48\hsize}
  \begin{center}
   \includegraphics[width=6.5cm]{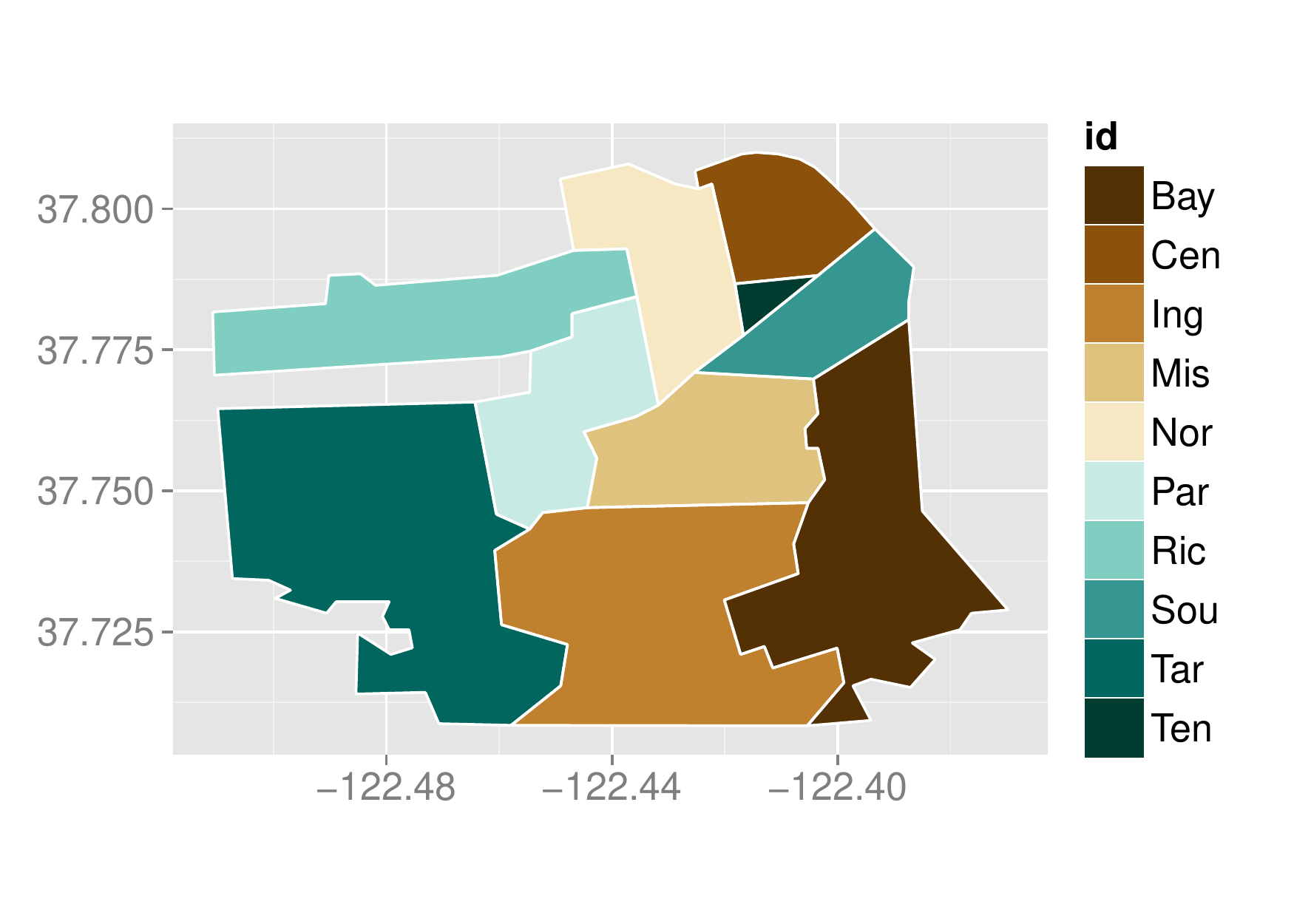}
  \end{center}
  \label{fig:SF}
 \end{minipage}
 \hfill
 \hfill
 \begin{minipage}{0.48\hsize}
  \begin{center}
   \includegraphics[width=6.5cm]{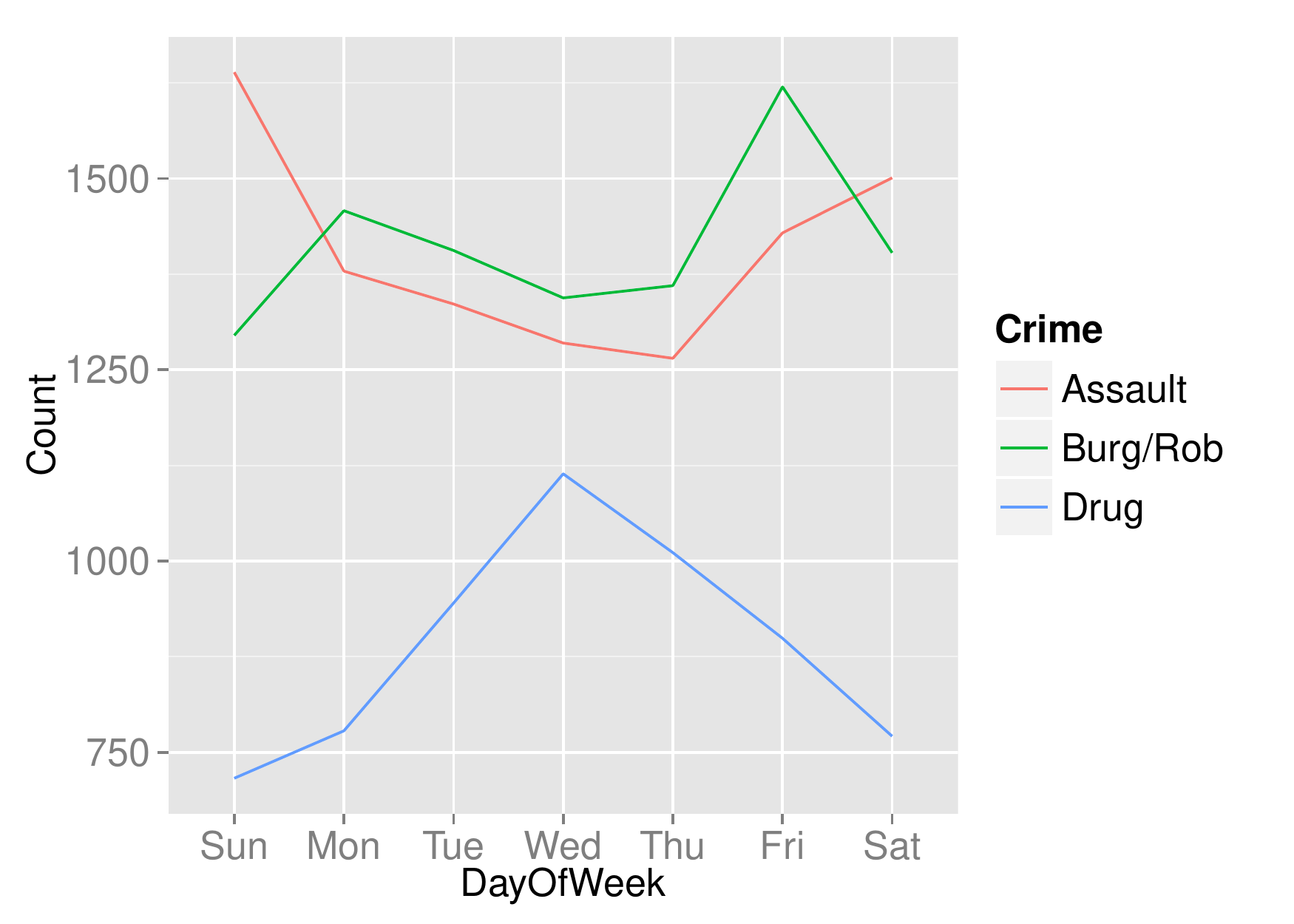}
  \end{center}
  \label{fig:Hist}
\end{minipage}
\end{figure}

Our dataset consists of crime events in the city of San Francisco in 2012. We have three crime type categories: (1) assault, (2) burglary/robbery\footnote{In the original dataset, burglary and robbery events are reported separately with hourly histograms and spatial density maps provided in the supplementary materials.  Burglary and robbery are not universally accepted as being behaviorally similar. However, we aggregate these crimes due to their similar definition and to increase the number of events in our point patterns.} and (3) drug. Each crime event has a time (date, day of week, time of day) and location (latitude and longitude) information.  Spatial coordinates (latitude and longitude) were transformed into eastings and northings.
Each crime event is also classified into a district.  In particular, there are 10 districts in San Francisco: (1) Bayview, (2) Central, (3) Ingleside, (4) Mission, (5) Northern, (6) Park, (7) Richmond, (8) Southern, (9 ) Taraval, (10) Tenderloin (see Figure \ref{fig:SF}).

Figure \ref{fig:Hist} shows the counts of crime events for day of week\footnote{Here, and in the sequel, we take day of the week as 02:00 to 02:00. This definition interprets crime events on, e.g., Saturday night as including the early hours of Sunday morning.}.  Counts for crime types show different patterns. Assault events happen more on weekends, but burglary/robbery events happen most on Friday. Interestingly, it seems drug events happen most on Wednesday.

\begin{figure}[htbp]
  \caption{Histograms of crime events by type and by day of the week}
  \begin{center}
   \includegraphics[width=13cm]{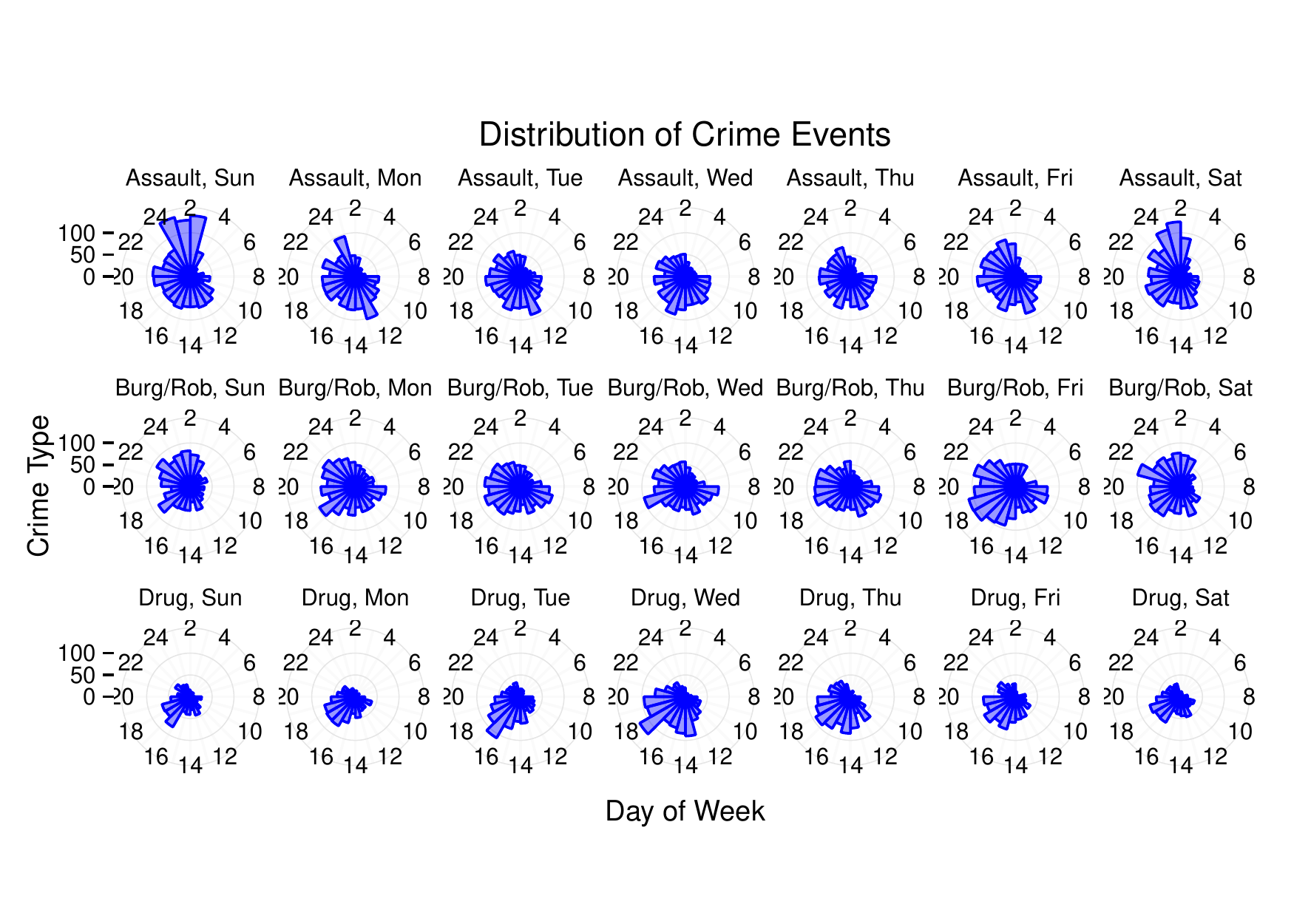}

  \end{center}
  \label{fig:RD_2012}
\end{figure}

Figure \ref{fig:RD_2012} shows the data by type and by day of the week ($3\times 7$ plots) in the form of `rose' diagrams. This figure reveals differences among crime types and also differences across day of the week. For example, drug-related crime events are observed more from 5 to 7 pm. while burglary/robbery crime events are observed later in the day.
Overall, the circular time dependence of crime events is seen, i.e., large counts from evening to late night and small counts from early morning through the middle of the day. In the point pattern model construction below, we model each crime type separately and, within crime type, incorporate day of week as a mark.

\section{Modeling and Theory}
Observations on a circle lead us to the world of directional data, as illustrated in Figure \ref{fig:RD_2012}.
Once an orientation has been chosen, the circular observations are specified using the angle from the orientation to the corresponding point on the unit circle. However, here, we are only concerned with point patterns on a circle.  For the nonhomogeneous Poisson process and log Gaussian Cox process models we only need to specify the intensity functions for the processes.  So, in what follows, we only consider specifications for intensities for space-time point patterns over $D \times S^1$ where $S^1$ is the unit circle.
\subsection{The nonhomogeneous Poisson process (NHPP) and Log Gaussian Cox process (LGCP)}

Again, since the crime events are random both in number and in space-time location, it is natural to think of them as a random point pattern over space and time.  Here, we make the assumption that events are located in space and time, conditionally independent given their intensity, anticipating that the intensity surface will explain the observed clustering in space and time.  So, we consider the two most common models for such a setting: the NHPP and the LGCP.
The LGCP dates at least to \cite*{Molleretal(98)}. As a spatial process, it is defined so that the log of the intensity is a Gaussian process (GP), i.e.,

\begin{equation}
\log \lambda(\bm{s})=X(\bm{s})^{T}\beta+Z(\bm{s}), \quad Z(\bm{s})\sim \mathcal{GP}(\bm{0}, C).
\end{equation}
Here, $Z(\bm{s})$ is a zero mean stationary, isotropic GP over $D$ with covariance function $C$, which provides spatial random effects for the intensity surface, pushing up and pulling down the surface, as appropriate.  If we remove $Z(\bm{s})$ from the log intensity, we obtain the associated NHPP.  NHPP's have a long history in the literature (see, e.g., \cite{Illianetal(08)}).  In fact,
if $Y$ is a Cox process with intensity $\Lambda(\bm{s})$, then, conditional on $\Lambda(\bm{s})=\lambda(\bm{s})$, $Y$ is an NHPP with intensity $\lambda(\bm{s})$.
Evidently, an LGCP provides a very flexible intensity specification.  Below, we will argue that, with regard to our crime data, we prefer the additional flexibility of a \emph{space-time} Log Gaussian Cox Process (LGCP) to the associated NHPP.  Our model is in the spirit of the space time LGCP introduced in \cite{BrixDiggle(01)}.
\subsection{Circular covariance functions for Gaussian processes}

Again, we consider a three dimensional Gaussian process with a two dimensional location, and one dimensional circular time.  In general, we seek

\begin{align}
Z(\bm{s},t)\sim \mathcal{GP}(0, C), \quad (\bm{s}, t)\in \mathbb{R}^2\times S^1
\end{align}
We need to specify valid correlation functions over $\mathbb{R}^{2} \times S^{1}$.




\cite{Gneiting(13)} proposes families of circular correlation functions (CCF's) based on truncation of familiar spatial correlation functions. He shows that the completely monotone functions  are strictly positive definite on spheres of any dimension, e.g., powered exponential, Mat\'{e}rn, generalized Cauchy, and Dagum families. One of the examples in \cite{Gneiting(13)} is the powered exponential family,
\begin{align}
C_{PE}(u)=\exp\biggl(-(\phi u)^{\alpha}\biggl), \quad u\in [0,\pi], \quad \alpha\in (0,1].
\end{align}
If $\alpha\in(0,1]$, this function is strictly positive definite function for any dimension, but if $\alpha>1$, then (5) is no longer positive definite, even in one dimension.

Another example in \cite{Gneiting(13)} is the generalized Cauchy family,
\begin{align}
C_{GC}(u)&=\biggl(1+(\phi u)^{\alpha}\biggl)^{-\tau/\alpha}, \quad \text{for} \quad u\in[0, \pi] \quad \alpha\in(0,1], \quad \tau>0.
\label{eq6}
\end{align}
where $\tau$ is a shape parameter which doesn't affect the positive definiteness as long as $\tau>0$.
This function is positive definite for any dimension if $\alpha \in(0,1]$. Again,  for $\alpha>1$, (\ref{eq6}) is also not positive definite, even in one dimension.

It may be surprising that restriction of familiar spatial correlation functions to the spherical domain maintains positive definiteness on the sphere. However, this enables convenient choices and, in fact, we adopt the generalized Cauchy family as the circular correlation function in the analysis below.

\subsection{Space and linear time covariance functions}
Next, we turn to valid covariance functions over $R^2 \times S^1$.  We consider both the separable case, which is immediate, and also the nonseparable case.\\

\noindent{\bf Separable covariance functions}\\

In the context of the LGCP model, we need to specify the covariance function for the latent Gaussian process $Z(\bm{s},t)$. Separable space time covariance functions are often adopted due to convenient specification and computational simplification {\citep{BanerjeeCarlinGelfand(14)}}. The separable specification arises if the space time covariance function is written as a product of a valid space and a valid time covariance function, i.e.,
\begin{align}
C_{s,t}(\bm{h},u)=C_{s}(\bm{h})C_{t}(u)
\end{align}
In our setting, we can define a valid space-time covariance function merely by choosing as $C_{s}$ any valid covariance function on $R^2$ and multiplying it by any of the foregoing valid CCF's.  The resulting covariance matrix for a set of $(\bm{s},t)$'s with $N$ $\bm{s}$'s by  $M$ $t$'s will have a Kronecker product form $C_{s}\otimes C_{t}$ where $C_{s}$ and $C_{t}$ are $N\times N$ and $M\times M$ covariance matrices.  Simplified inverse, determinant, and Cholesky decomposition result,
making  the separable specification computationally efficient and tractable in high dimensional cases.

In this regard, we note that the point pattern data arises as a set $(\bm{s}_{i},t_{i}), i=1,2,...,n$ where $n$ is the total number of points.  We don't have a factorization in space and time so why is the separable form helpful?  Below, we clarify the need for grid approximation (discretization) for both space and time in order to evaluate the LGCP likelihood and in order to obtain manageable computation for the model fitting. Then, $N$ and $M$ will become the number of grid centroids for space and for time respectively. For the separable case, we can then take advantage of the Kronecker factorization. For the nonseparable case, we require the Cholesky decomposition and the inverse of an $NM\times NM$ matrix, making the computation much more demanding.\\

\noindent{\bf Nonseparable covariance functions}\\

It is evident that the separable covariance specification is restrictive for real data applications because it precludes space-time interaction of the sort we mentioned in the Introduction.
Various versions of nonseparable covariance functions have been proposed for the case where space is again $R^2$ and time is linear. {\cite{CressieHuang(99)} propose specifications of nonseparable stationary covariance functions based on Fourier transformation on $\mathbb{R}^{d}$ with criteria which guarantee positive definiteness.
Since their specification requires closed form solution for the $d$ dimensional Fourier transformation, the class of functions is relatively small.
\cite{Gneiting(02)} proposed a flexible parametric family of nonseparable covariance functions, extending the results of \cite{CressieHuang(99)}.
Gneiting's class takes the form,
\begin{align}
C(\bm{h},u)=\frac{\sigma^2}{\psi(\|u\|^{2})^{d/2}}\varphi\biggl(\frac{\|\bm{h}\|}{\psi(\|u\|^2)}\biggl), \quad (\bm{h}, u)\in \mathbb{R}^{d}\times \mathbb{R}^{l}
\end{align}
where $\varphi(t)$, $t\ge 0$, is a completely monotone function and $\psi(t)$, $t\ge 0$, is a positive function with a completely monotone derivative.
In our modeling below, we utilize Gneiting's specification. However, these covariance functions are specified on $\mathbb{R}^{d}\times \mathbb{R}^{1}$; our need is to provide valid nonseparable covariance functions on $\mathbb{R}^{d}\times S^1$.

\subsection{Space by circular time covariance functions}

Finally, we turn to the space-time covariance functions we seek.  For the spatial correlation function, we assume the exponential correlation function,
\begin{align}
C_{s}(h)=\exp\biggl(-\phi_{s}\|h\| \biggl).
\end{align}
For the circular time correlation function, we consider the truncated generalized Cauchy correlation function (again, see, \cite{Gneiting(13)}).



So, we arrive at the following proposed nonseparable covariance function over space by circular time:
\begin{align}
C(\bm{h},u)&=\frac{\sigma^2}{(1+(\phi_{t}u)^{\alpha})^{\delta+\gamma(d/2)}}\exp\biggl(-\frac{\phi_{s}\|\bm{h}\|}{(1+(\phi_{t}u)^{\alpha})^{\gamma/2}}\biggl), \\
\gamma &\in (0,1] \quad \alpha \in (0,1]  \nonumber
\end{align}
where $(\bm{h},u)\in \mathbb{R}^2\times [0,\pi]$ and $\gamma$ is the nonseparability parameter. We can show that this is a valid covariance function on   $\mathbb{R}^{2}\times S^{1}$ following the proof in \cite{Gneiting(02)}, working with the valid circular choices according to \cite{Gneiting(13)}. We prove this result in the supplemental material.

As noted in the Introduction, motivated by processes observed on the Earth over time, recently, \cite{Porcuetal(15)} proposed nonseparable covariance functions on spheres crossed with linear time (as well as cross-covariance functions for multivariate random fields defined over a sphere). Hence, in space and time, their nonseparable covariance functions are specified on $S^{d}\times \mathbb{R}$ where $S^d$ is the $d$-dimensional unit sphere.
One of their contributions is similar to ours, i.e., following work by \cite{Gneiting(02)} presenting nonseparable space-time covariance functions with further work by Gneiting \cite{Gneiting(13)} presenting spatial correlation functions on a sphere, they obtain fairly general nonseparable forms (see Table 2 in online supplementary material of \cite{Porcuetal(15)}) over this product space.  Here, we follow a similar path but take space as $R^2$ with time as circular and, for our application, we employ the particular class of the generalized Cauchy family over this product space. As a result, the case of the real line crossed with the circle provides the common domain.

For model fitting with (3.8), we need to implement calculations for an $NM\times NM$ matrix.  For the whole of the city of San Francisco, $N$ is very large. Thus, for convenience, we take a smaller region and adopt a more local investigation of nonseparability for space-time crime patterns.  Figure \ref{fig:Point} shows the spatial locations of all of the crime events and the point patterns for the Tenderloin and Mission districts where relatively more events are observed than in the other districts. So, we create a rectangular region around this area (see Figure \ref{fig:Land} below).  For this region, in the interest of comparison, we implement the truncated generalized Cauchy correlation function for circular time in both the separable and nonseparable cases.

\begin{figure}[htbp]
  \caption{Crime event locations in SF (upper), Tenderloin (middle) and Mission (lower) for assault (left), burglary/robbery (middle) and drug (right)}
 \begin{minipage}{0.32\hsize}
  \begin{center}
   \includegraphics[width=4.5cm]{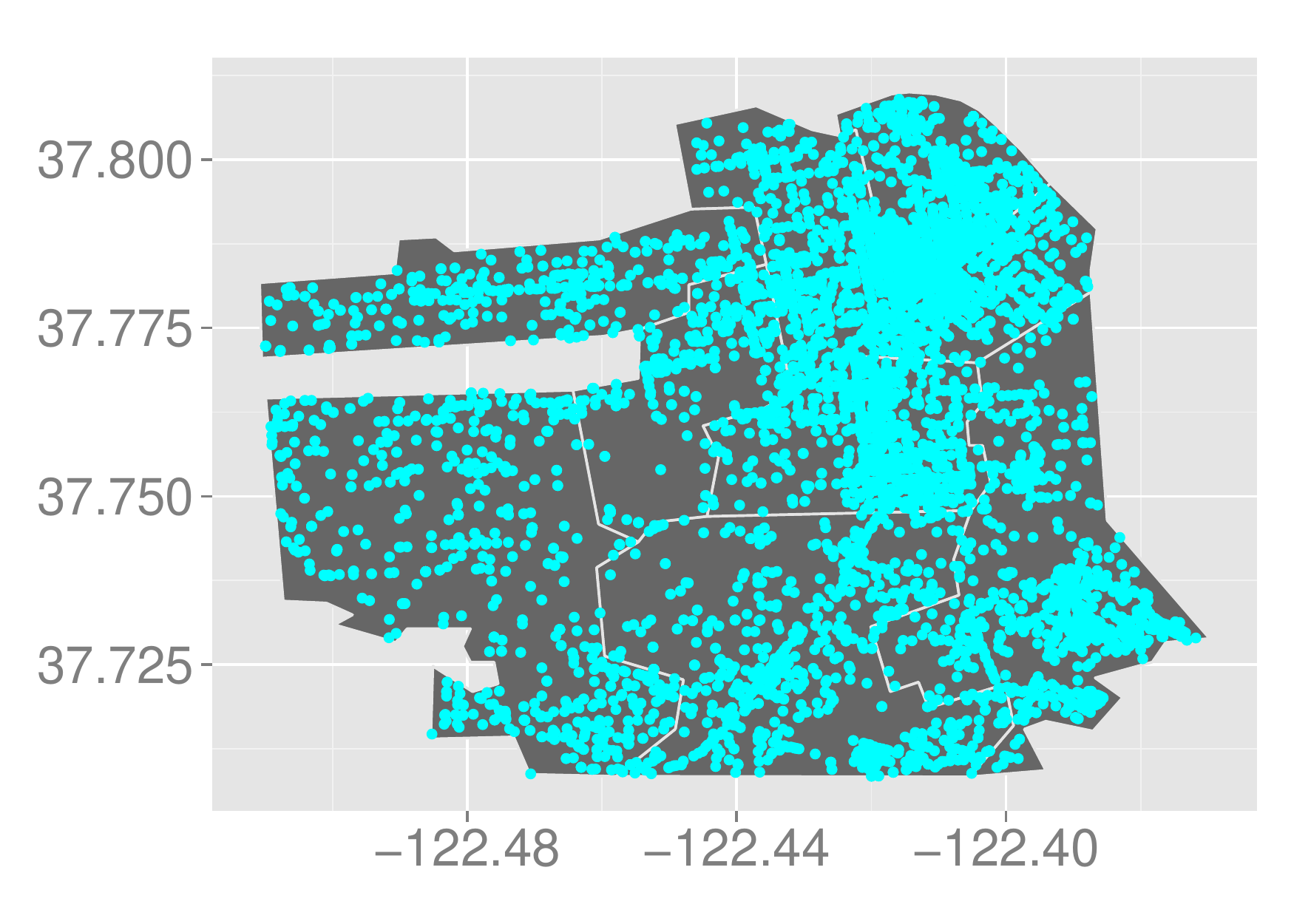}
  \end{center}
 \end{minipage}
 \hfill
 \begin{minipage}{0.32\hsize}
  \begin{center}
   \includegraphics[width=4.5cm]{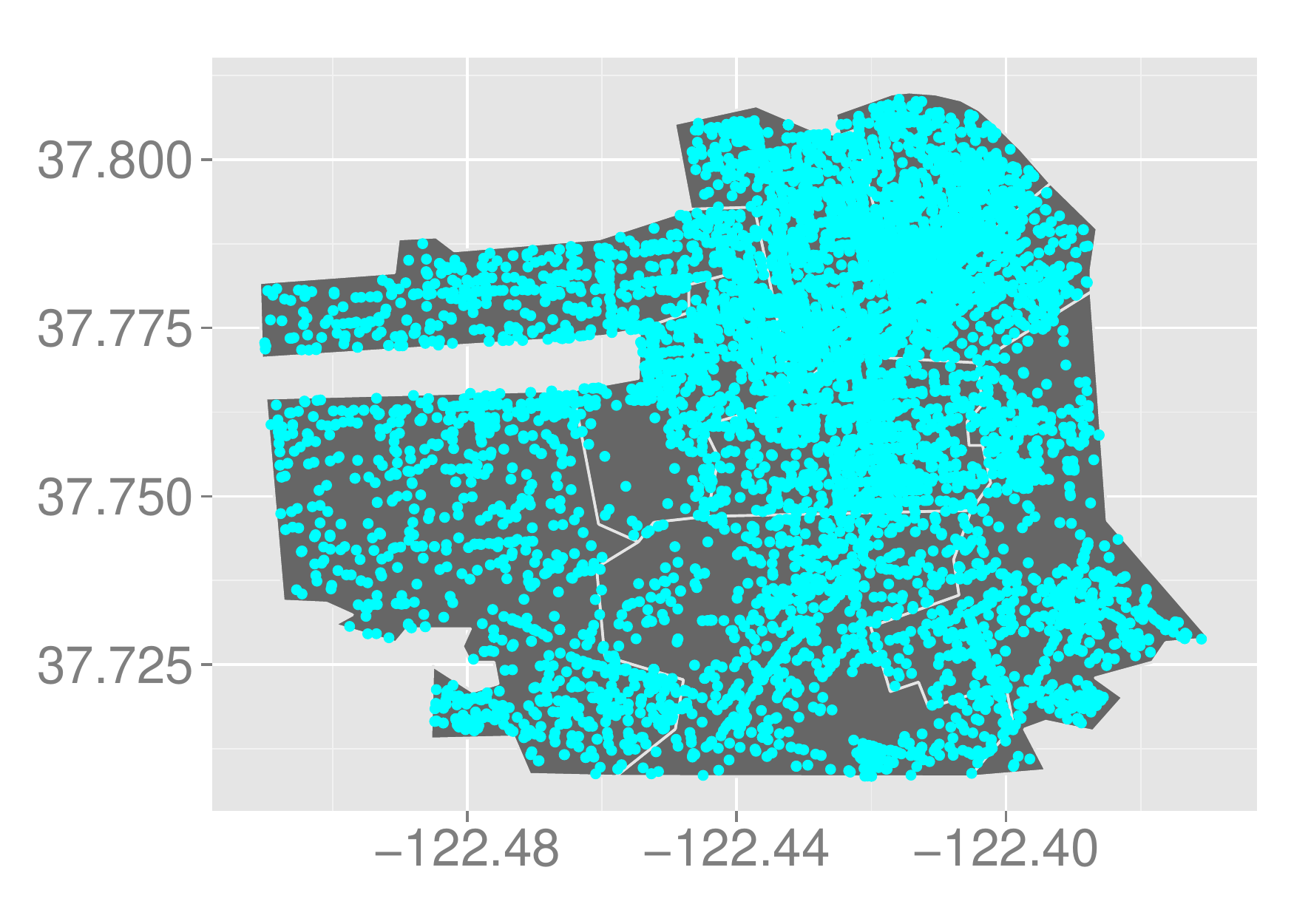}
  \end{center}
 \end{minipage}
 \hfill
 \begin{minipage}{0.32\hsize}
  \begin{center}
   \includegraphics[width=4.5cm]{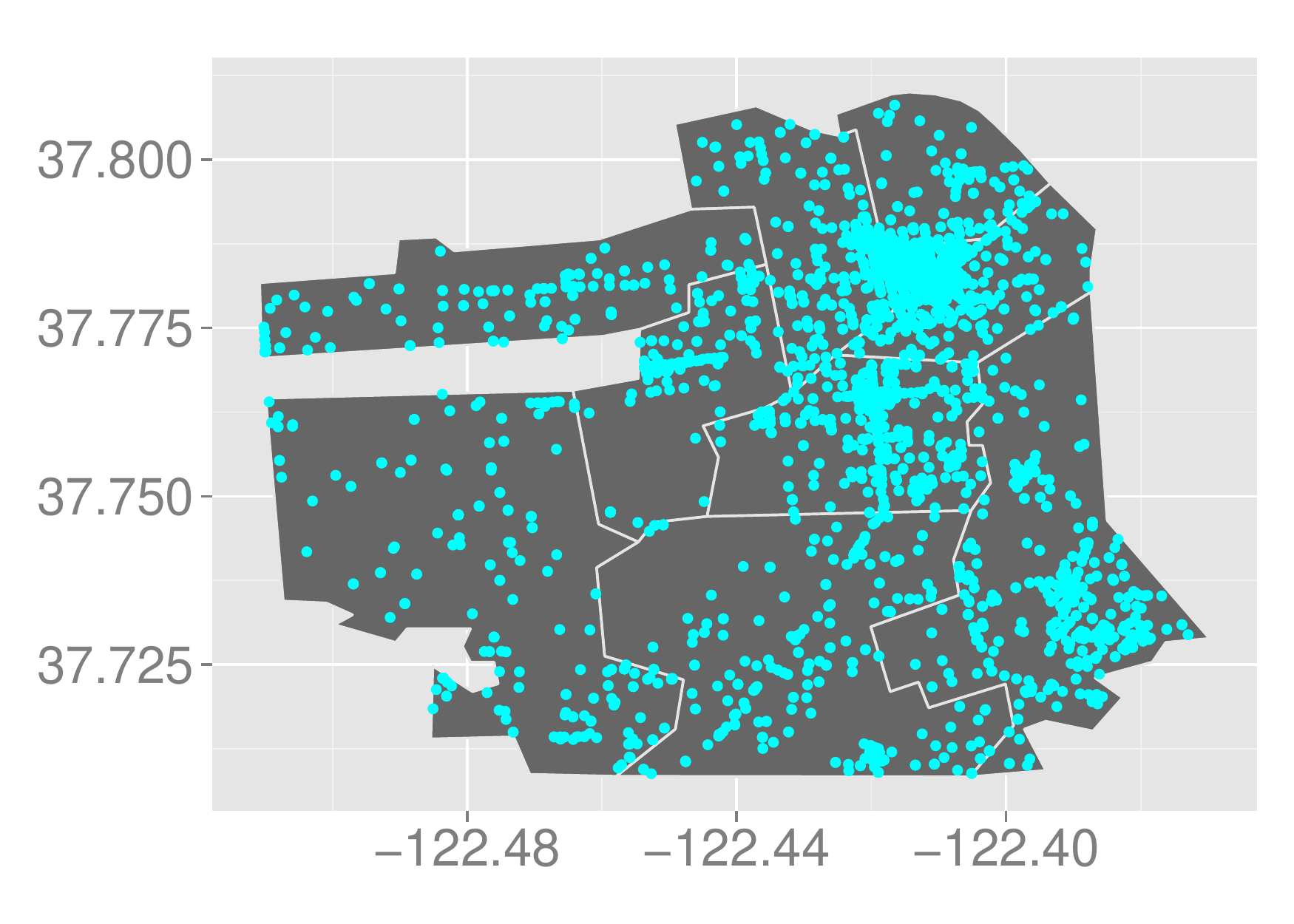}
  \end{center}
 \end{minipage}
 \begin{minipage}{0.32\hsize}
  \begin{center}
   \includegraphics[width=4.5cm]{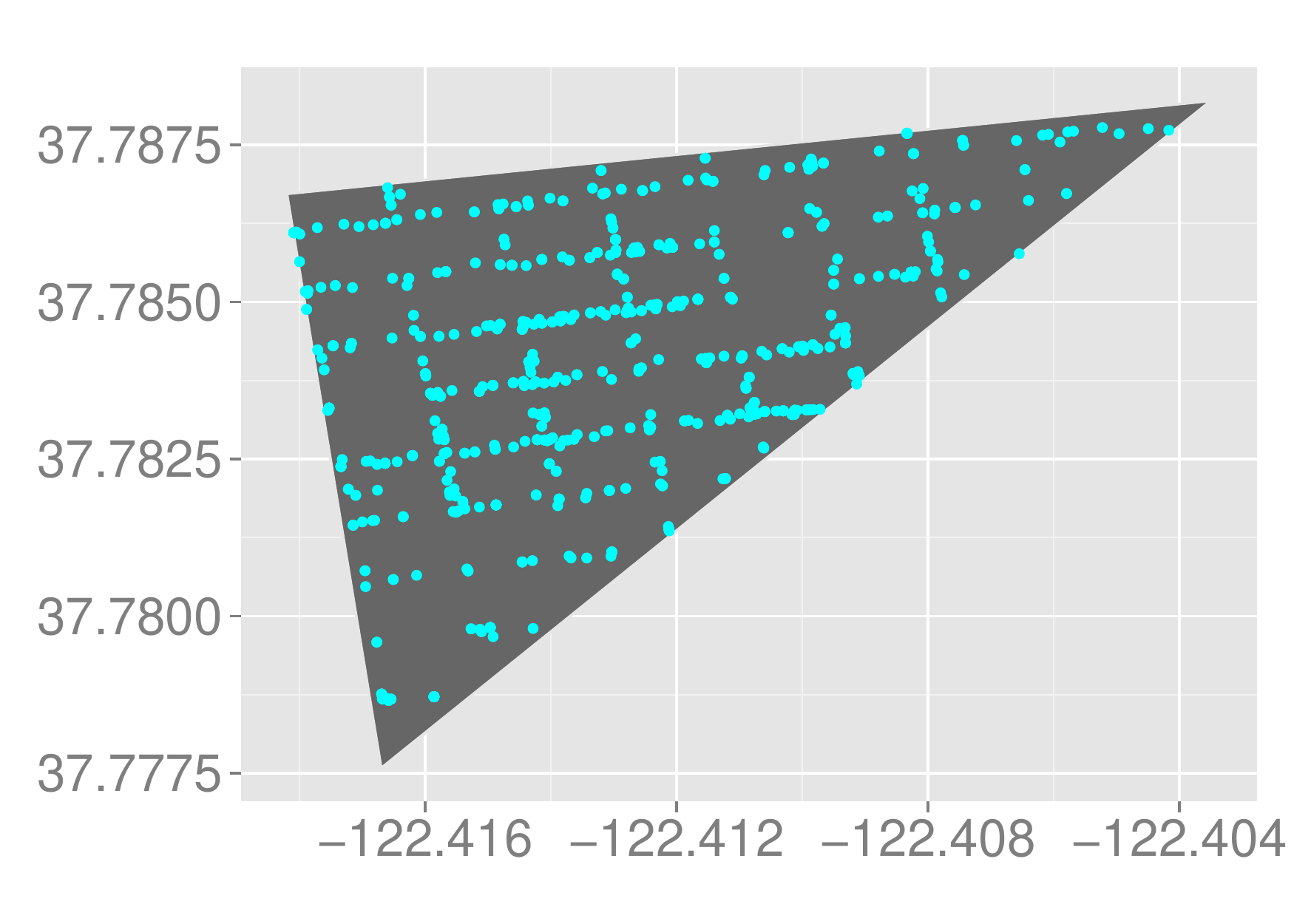}
  \end{center}
 \end{minipage}
 \hfill
 \begin{minipage}{0.32\hsize}
  \begin{center}
   \includegraphics[width=4.5cm]{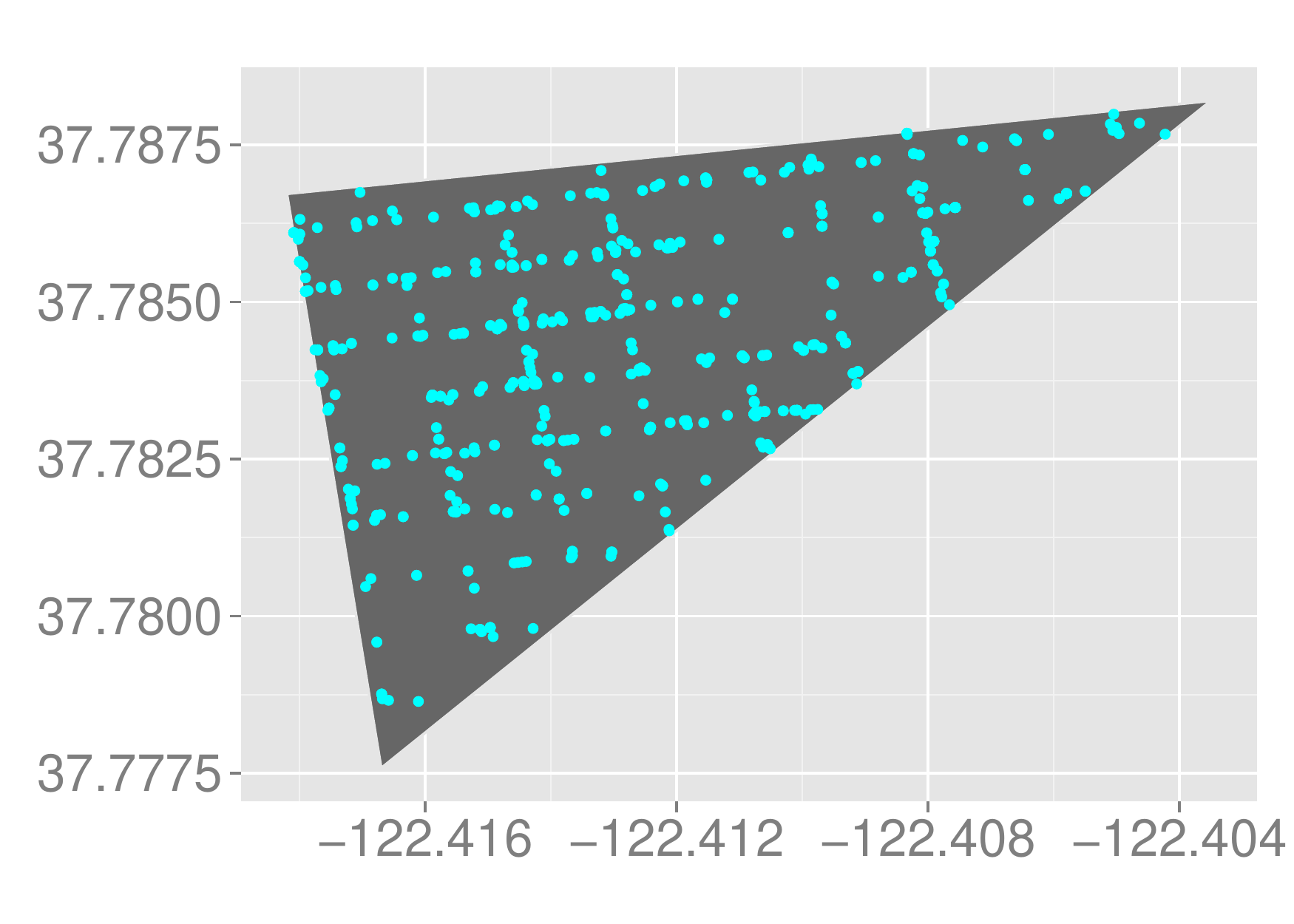}
  \end{center}
 \end{minipage}
 \hfill
 \begin{minipage}{0.32\hsize}
  \begin{center}
   \includegraphics[width=4.5cm]{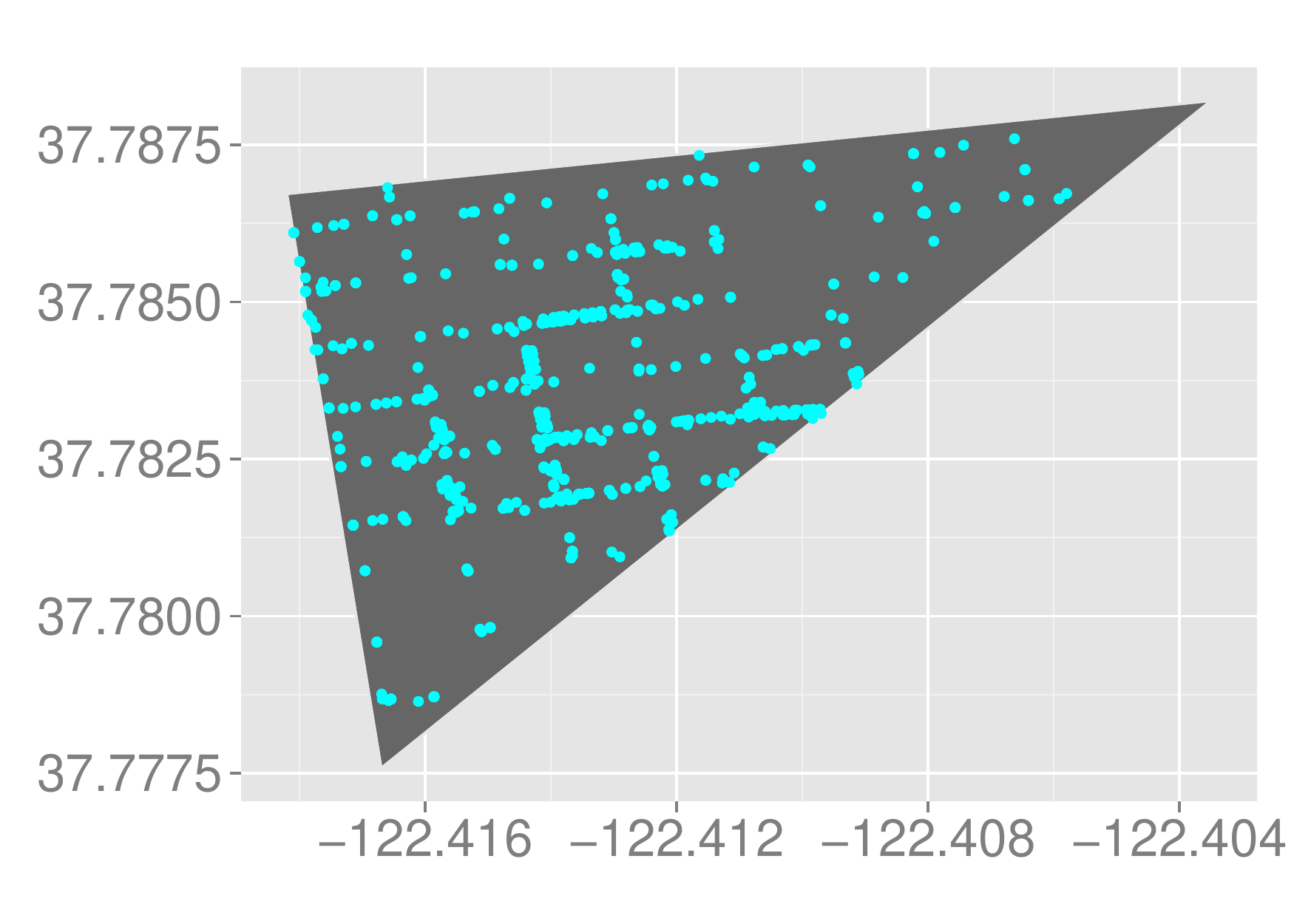}
  \end{center}
 \end{minipage}
 \begin{minipage}{0.32\hsize}
  \begin{center}
   \includegraphics[width=4.5cm]{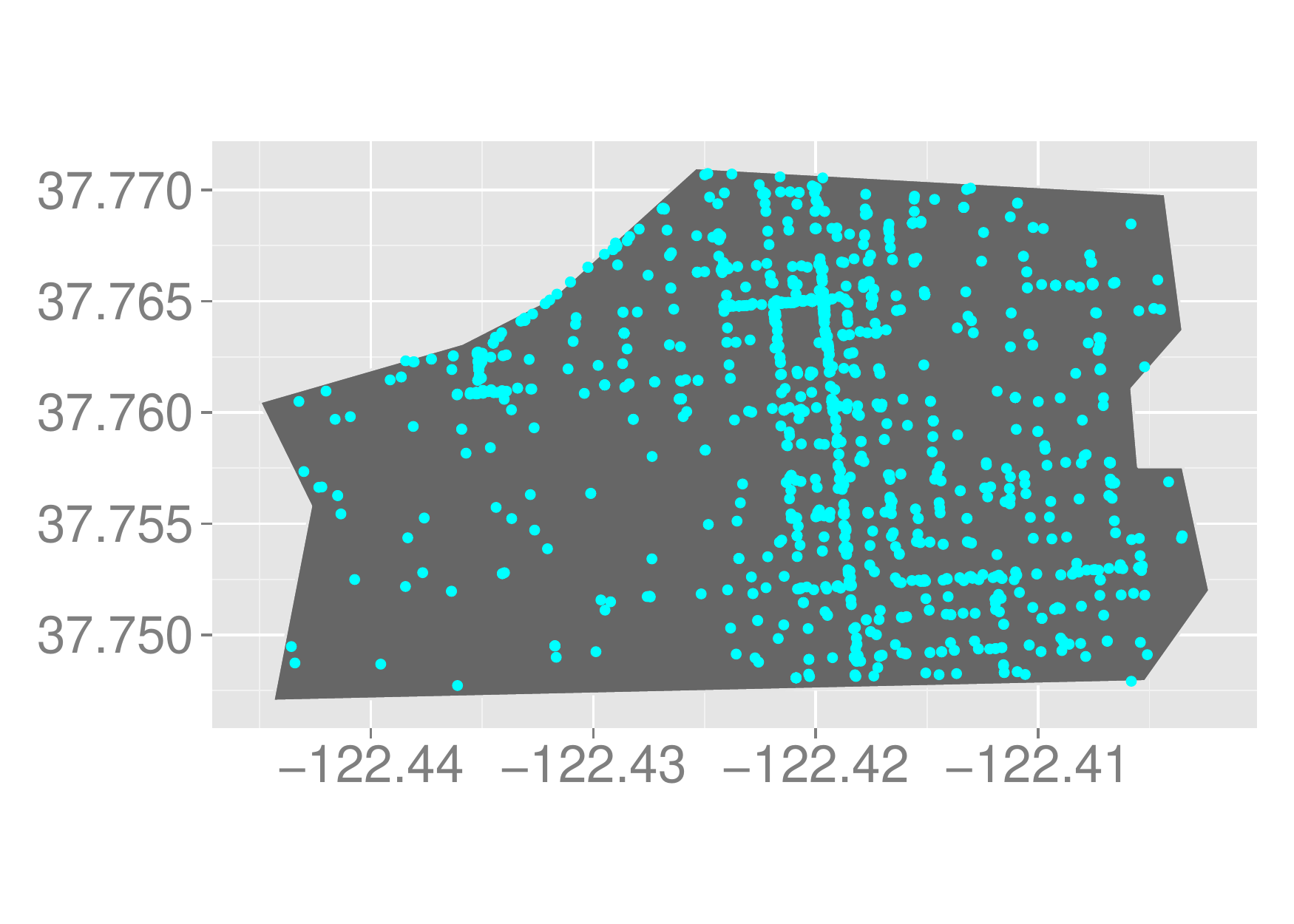}
  \end{center}
 \end{minipage}
 \hfill
 \begin{minipage}{0.32\hsize}
  \begin{center}
   \includegraphics[width=4.5cm]{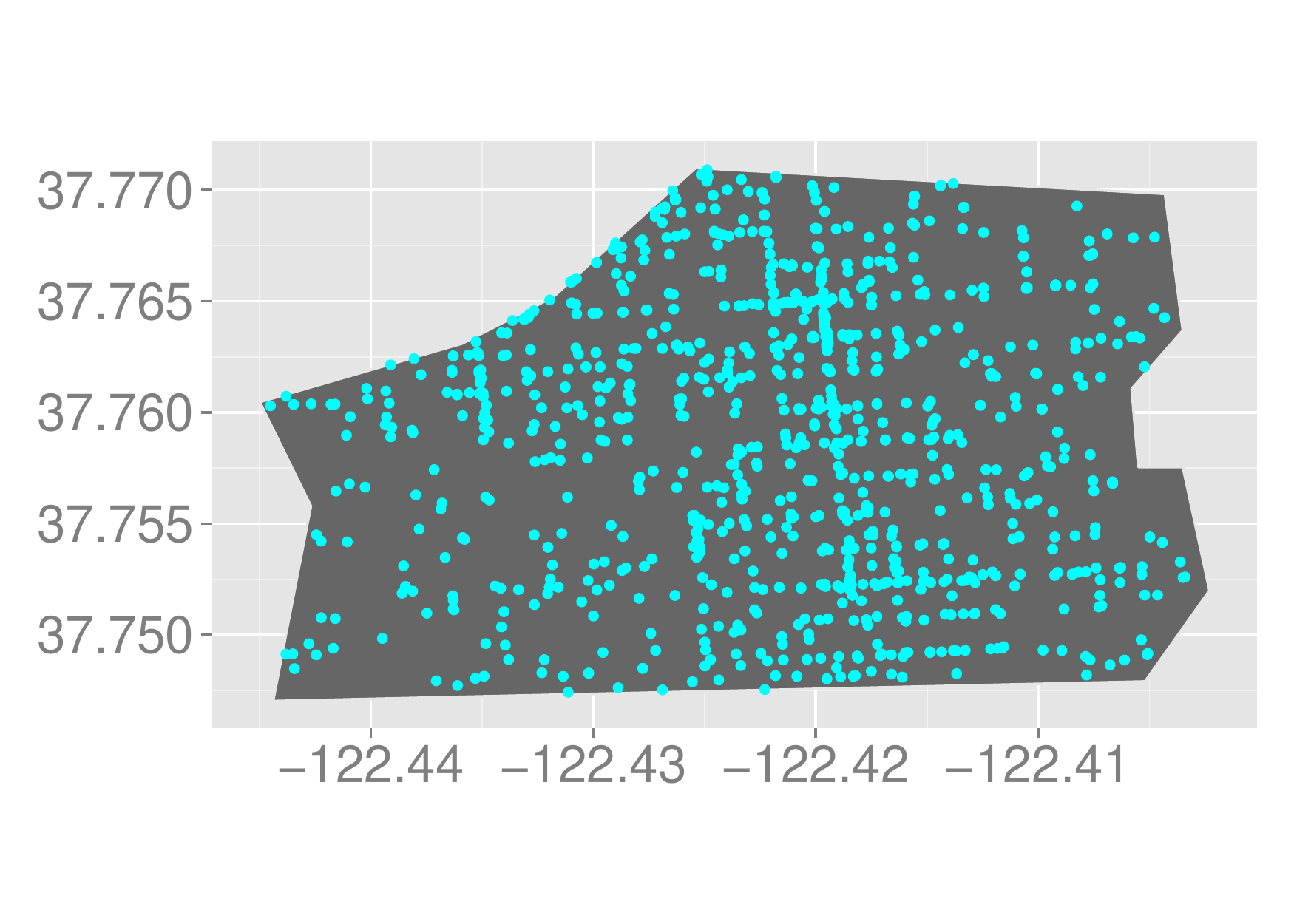}
  \end{center}
 \end{minipage}
 \hfill
 \begin{minipage}{0.32\hsize}
  \begin{center}
   \includegraphics[width=4.5cm]{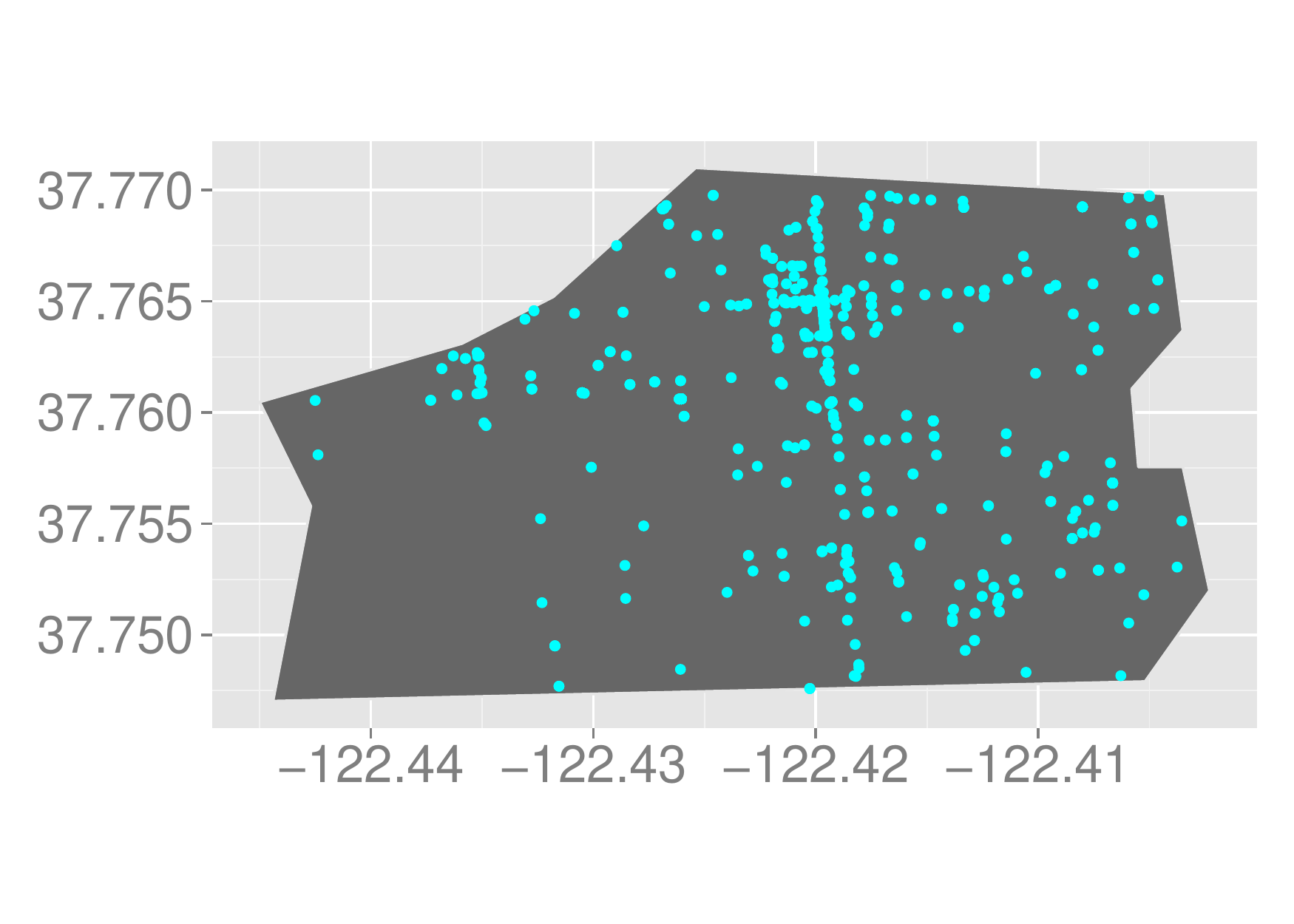}
  \end{center}
 \end{minipage}
  \label{fig:Point}
\end{figure}

\section{Model specification, fitting and checking}
We specify the intensity for the NHPP as $\log \lambda(\bm{s},t) = \log \lambda_{0}(\bm{s})+ \log \kappa_{0}(t)$.  For the LGCP, we add $Z(\bm{s},t)$, a mean $0$ GP with covariance function chosen according to the previous section. \\

\noindent{\bf Covariate specification}\\

We employ constructed space and time covariates.
For the spatial covariates, we identify a set of landmarks.
These landmarks are referred to as \emph{crime attractors} (\cite{Brantingham(95)}) and are  selected from centers of commercial activity, i.e., places with high population density, high human exposure. Examples might include malls, market streets, and amusement centers.
For a given landmark, we employ a directional Gaussian kernel function as the distance measure from crime location to landmark. That is, inverse distance measures risk; the smaller the distance the larger the risk.

Formally, the covariate level at location $\bm{s}$ associated with landmark $k$ is
\begin{align}
g_{k}(\bm{s}, \bm{s}^{*}_{k})=\exp\biggl(-\frac{1}{2}(\bm{s}-\bm{s}^{*}_{k})^{'}\Sigma_{k}^{-1}(\bm{s}-\bm{s}^{*}_{k})\biggl)
\end{align}
where $\bm{s}^{*}_{k}$ is the location of landmark $k$ and $\Sigma_{k} = \left(
                                                                           \begin{array}{cc}
                                                                             \sigma_{1}^{2} &
                                                                             \rho_{k}\sigma_{1}\sigma_{2} \\
                                                                             \rho_{k}\sigma_{1}\sigma_{2} &
                                                                             \sigma_{2}^{2} \\
                                                                           \end{array}
                                                                         \right)$
is a positive definite matrix for landmark $k$.
Here, $\sigma_{1}^2$ and $\sigma_{2}^2$ are scales for Easting and Northing coordinates and $\rho_{k}$ is the correlation of the kernel for landmark $k$.  In fact, $\sigma_{1}$ is the centroid to centroid Easting distance between adjacent grid cells, $\sigma_{2}$ the centroid to centroid Northing distance between adjacent grid cells.  $\rho_{k}$ is treated as an unknown, along with the $\beta_{k}$'s, a regression coefficient assigned to $g_{k}$.
Thus, we let $\log\lambda_{0}(\bm{s})=\sum_{k=1}^{K}\beta_{k}g_{k}(\bm{s}, \bm{s}^{*}_{k})$.

Figure \ref{fig:Land} shows the contour plot for assault events and the landmarks. Illustratively, we create two landmarks, $L_{1}=(-122.408, 37.784)$ (Union Square Shopping Center, henceforth ``Union Square'') and $ L_{2}=(-122.419, 37.764)$ (BART Station, 16th and Mission, henceforth ``BART Station''). The left hand side is the contour plot of the kernel density estimate for the observed assault events, obtained by \texttt{ggplot2} in R package.  The right hand side of figure \ref{fig:Land} overlays the subregion used with the nonseparable covariance.

\begin{figure}[htbp]
  \caption{The contour plot of the kernel density estimate for the assault events (left) and landmarks with the subregion (right, color rectangular region)}
 \begin{minipage}{0.48\hsize}
  \begin{center}
   \includegraphics[width=6.5cm]{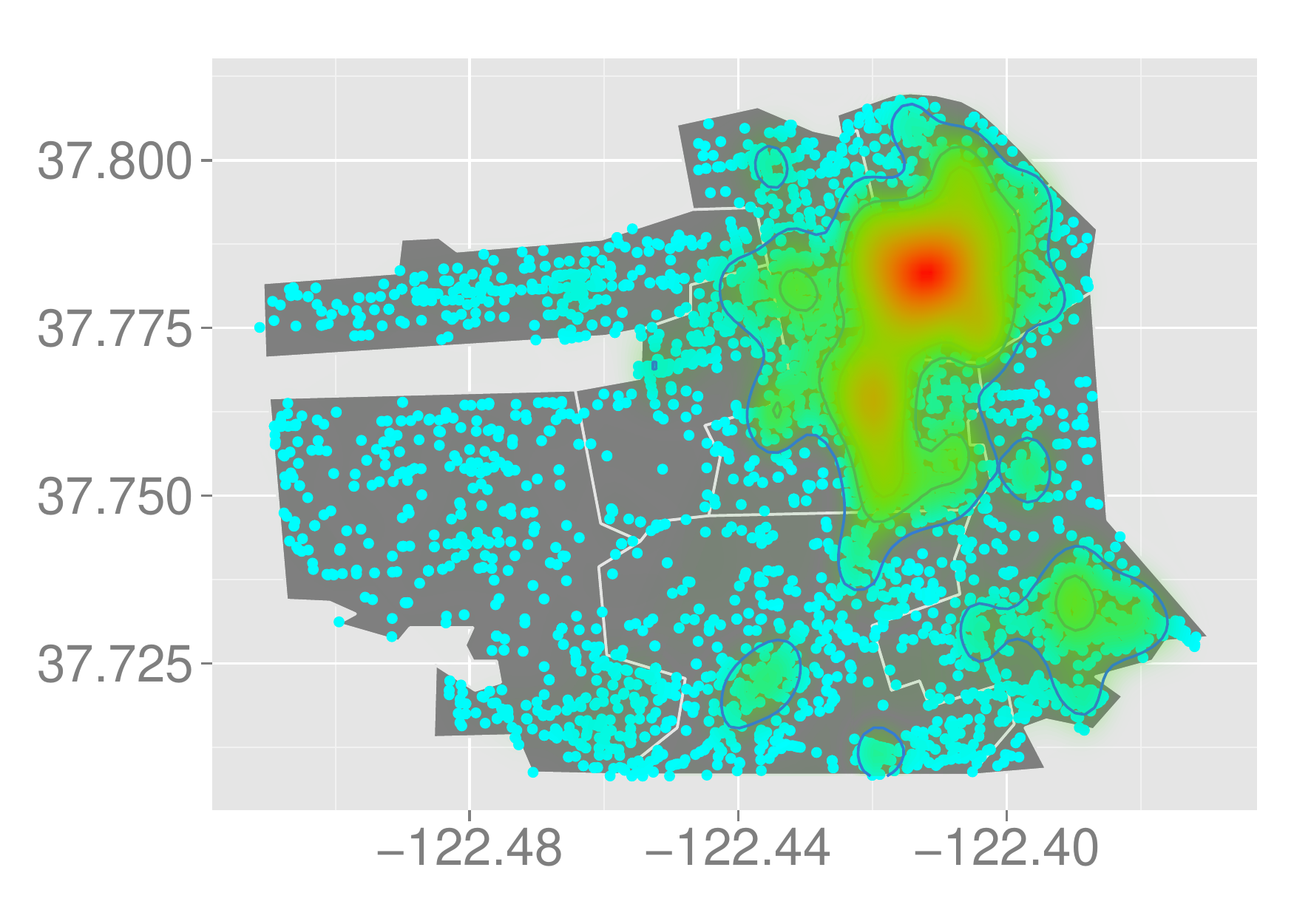}
  \end{center}
 \end{minipage}
 \hfill
 \begin{minipage}{0.48\hsize}
  \begin{center}
   \includegraphics[width=6.5cm]{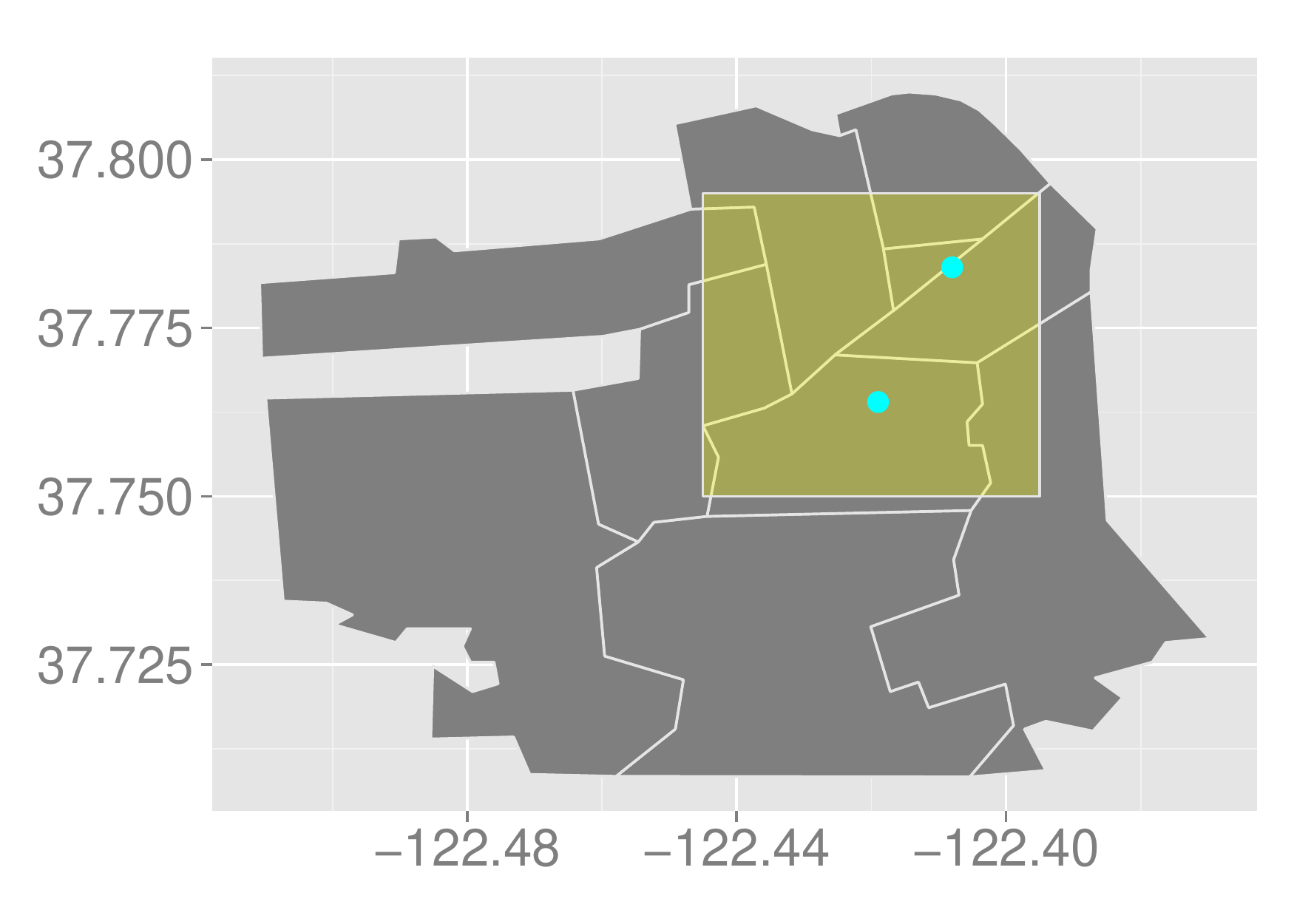}
  \end{center}
 \end{minipage}
  \label{fig:Land}
\end{figure}

To form a temporal covariate we need a function whose support is the unit circle.  Since crime events occur more frequently in the evening and night hours, less in the morning and afternoon hours, the most elementary constructed covariate which reflects this would have two levels.  Here, we let
\begin{equation}
\kappa(t) = \mu(1 + \delta 1(t \in [4\pi/3, 2\pi))).
\end{equation}
On the $24$ hour scale, this choice of $\kappa$ would be interpreted as adopting level $\mu$ for times between 02:00 and 18:00 and level $\mu(1 + \delta)$ for times between 18:00 and 02:00 in the morning. $\mu$ and $\delta$ become model parameters; alternative windows could be explored. In fact, as demonstrated in Figure \ref{fig:RD_2012}, the pattern and incidence of crime events vary across day of week. So, we introduce a different temporal covariate for each day of the week (writing $\mu_{w}$ and $\delta_{w}$ for $w=Sun,\ldots,Sat$).  Recalling Section 2, we make these covariates consistent by defining day of the week as 02:00 to 02:00.  This specification yields $\kappa(t,w)$ of the form in (4.2).  Combined with $g_{k}(\bm{s}, \bm{s}^{*}_{k})$ in (4.1), this enables our specification of $\lambda(\bm{s}, t, w)$ below.\\

\noindent{\bf Model specification}\\

Our baseline space by circular time LGCP model with \emph{separable} space-time covariance function is defined below.
We employ the foregoing binary time of day covariate and landmark distance covariates.
The model is defined on grid points and on each day of week. After discretization, let $J$ be the total number of space-time grids cells, i.e., we have $(\bm{s}_{j}, t_{j})$, for $j=1,\ldots, J$ and $w=Sun,\ldots,Sat$
\begin{align}
y_{\bm{s}_{j}, t_{j}, w}|\lambda(\bm{s}_{j},t_{j},w)&\sim \mathcal{NHPP}(\lambda(\bm{s}_{j},t_{j},w)), \quad \bm{s}_{j}\in \mathbb{R}^2 \quad t_{j}\in S^{1} \\
\lambda(\bm{s}_{j},t_{j},w)&=\lambda_{0}(\bm{s}_{j})\kappa(t_{j},w)\exp\biggl(-\frac{\sigma^2}{2}+Z(\bm{s}_{j},t_{j})\biggl), \\ Z(\bm{s}_{j},t_{j}) &\sim \mathcal{GP}(0, C), \quad C=\sigma^2C_{s}\otimes C_{t} \\
\log\lambda_{0}(\bm{s}_{j})&=\sum_{k=1}^{K}\beta_{k}g_{k}(\bm{s}_{j}, \bm{s}^{*}_{k}), \\
g_{k}(\bm{s}_{j}, \bm{s}^{*}_{k})&=\exp\biggl(-\frac{1}{2}(\bm{s}_{j}-\bm{s}_{k}^{*})'\Sigma_{k}^{-1}(\bm{s}_{j}-\bm{s}_{k}^{*}) \biggl) \\
\kappa(t_{j},w)&=\sum_{v=Sun}^{Sat}\bm{1}(w=v)\mu_{v}(1+\delta_{v} \bm{1} (t_{j} \in \bm{I}))
\end{align}
where $y_{\bm{s}_{j},t_{j},w}$ is the count for grid cell $j$ on day $w$, $t\in S^{1}$ is the circular time variable, $\bm{s}$ is the location coordinates and $\bm{I}=[4\pi/3,2\pi)$. The space and circular time Gaussian process $Z(\bm{s},t)$ is common across all the day of week, while the temporal covariate $\kappa(t,w)$ is dependent on the day of week.
\\

\noindent{\bf Prior specifications} \\

We assume gamma priors for the time scale parameters $\mu$ and $\delta$ for convenience and normal priors for $\bm{\beta}$'s with large variance.
As for the parameters of Gaussian processes, we assume uniform distributions for $\phi_{s}$ and $\phi_{t}$.  The range associated with $\phi_{s}$ is chosen such that the correlation between locations at the maximum distance for the study region is 0.05. The maximum circular distance in time is $\pi$ so we chose $\phi_{t}$ to provide correlation $0.05$ at that distance.  We used these priors in both the separable and nonseparable cases.

For the spatial Gaussian processes, $\phi_{s}$ and $\sigma^2$ are not identifiable {\citep{Zhang(04)}} so we need to adopt an informative prior distribution for one of them. Here, we are informative about $\phi$ and adopt an inverse gamma distribution for $\sigma^{2}$ with relatively large variance. Finally, we assume a uniform prior on the domain of definition for the \emph{separability} parameter $\gamma$.
\subsection{Model fitting}

In fitting of the LGCP model, we have a stochastic integral of the form $\int_{D\times S^1}\lambda(\bm{s}, t)d\bm{s}dt$ in the exponential of the likelihood. We use grid cell approximation for this integral as well as for the product term in the likelihood yielding
\begin{align}
L(\lambda(\cdot))\propto \exp\biggl(-\sum_{w=Sun}^{Sat}\sum_{j=1}^{J} \lambda(\bm{s}_{j}^{*},t_{j}^{*}, w)\Delta_{s,t,w} \biggl)\prod_{w=Sun}^{Sat}\prod_{j=1}^{J}\lambda^{n_{j,w}}(\bm{s}_{j}^{*}, t_{j}^{*}, w)
\label{eq:like}
\end{align}
where $n_{j,w}$ is the number of events in grid $j$ on day $w$, $J$ is the number of grid cells, $N$ is the total number of points in the point pattern and  $(\bm{s}_{j}^{*}, t_{j}^{*})$ are the centroids of the grids cells. Fitting this approximation is straightforward because we only require evaluation of $\lambda(\bm{s}_{j}^{*}, t_{j}^{*}, w)$  over the grid cells.

Sampling of parameters related to the NHPP model can be implemented through the Metroplis Hastings (MH) algorithm. However, for the LGCP model, sampling the large number of $Z$'s from the Gaussian process is difficult to implement efficiently with standard Gibbs sampling. The customary Metropolis-Hastings algorithm often gets stuck in local modes, so a more sophisticated MCMC algorithm is required. A now-common approach for the LGCP model is to utilize the Metropolis adjusted Langevin algorithm (MALA) (see, \cite*{Molleretal(98)} and \cite{GirolamiCalderhead(13)}).


Here, for the Gaussian process outputs and hyperparameters,  we use  elliptical slice sampling (\cite*{MurrayAdamsMacKay(10)} and \cite{MurrayAdams(10)}) as discussed in {\cite{LeiningerGelfand(16)}}. 
Our sampling algorithm is based on algorithms 1 and 2 in \cite{Leininger(14)}, p.50-51). 
Let $\bm{Z}$ denote the vector of Gaussian process variables we need to sample, with $\bm{Z}$ having covariance matrix $C_{\bm{\theta}}$.  Let $\bm{Z}=L_{\bm{\theta}}\bm{\nu}$ where $C_{\bm{\theta}}=L_{\bm{\theta}}L_{\bm{\theta}}^{'}$ and $\bm{\nu}\sim N(0, I)$.  We sample $\bm{\nu}^{*}=\bm{\nu} \cos(\omega)+\bm{\eta} \sin(\omega)$ where $\bm{\eta}\sim N(0,I)$ through the elliptical slice sampling algorithm. Given a sampled $\bm{\nu}$, we sample the hyperparameters $\bm{\theta}$ by proposing the $\bm{\theta}^{*}\sim q(\bm{\theta}^{*}|\bm{\theta})$ and $\bm{Z}^{*}=L_{\bm{\theta}^{*}}\bm{\nu}$ such that
\begin{align*}
u<\frac{L(\bm{Z}^{*})\pi(\bm{\theta}^{*})q(\bm{\theta}|\bm{\theta}^{*})}{L(\bm{Z})\pi(\bm{\theta})q(\bm{\theta}^{*}|\bm{\theta})}, \quad u\sim \mathcal{U}[0,1]
\end{align*}
where $L(\cdot)$ is the likelihood in (\ref{eq:like}). We use random walk Metropolis Hastings for proposing the candidates, adaptively tuning the variance of the proposal density (see, \cite{AndrieuThoms(08)}). In this algorithm, sampling the hyperparameters does not involve direct evaluation of the prior distribution of the Gaussian process ($N(\bm{Z}|\boldsymbol{0},C_{\bm{\theta}})$) due to the transformation of variables.
\subsection{Model adequacy and model comparison}

Cross validation is a standard approach for assessing model adequacy and is available for point pattern models with conditionally independent locations given the intensity, as for both the NHPP and LGCP (see, \cite{LeiningerGelfand(16)}).

We implement cross validation by obtaining a training (fitting) dataset and a testing (validation) using $p$-thinning as proposed by \cite{LeiningerGelfand(16)}. Let $p$ denote
the retention probability, i.e., we delete $\bm{s}_{i}\in \mathcal{S}$ with probability $1-p$. This produces a training point pattern $\mathcal{S}^{train}$ and test point pattern $\mathcal{S}^{test}$, which are independent, conditional on $\lambda(\bm{s})$.  In particular, $\mathcal{S}^{train}$ has intensity $\lambda(\bm{s})^{train}=p \lambda(\bm{s})$. We set $p=0.5$ and estimate $\lambda(\bm{s})^{train}$ $\bm{s}\in D$.  Then, we convert the posterior draws of $\lambda^{train}(\bm{s})$ into predictive draws of $\lambda^{test}(\bm{s})$ using $\lambda^{test}(\bm{s})=\frac{1-p}{p}\lambda^{train}(\bm{s})$.

Let $\{B_{k}\}$ be a collection of subsets of $D$.
For the choice of  $\{B_{k}\}$, \cite{LeiningerGelfand(16)} suggest to draw random subsets of the same size uniformly over $D$.  Specifically, for $q \in (0,1)$, if the area of  each $B_{k}$ is $q|D|$, then $q$ is the \emph{relative} size of each $B_{k}$.   They argue that making the subsets disjoint is time consuming and unnecessary.
Based on the $p$-thinning cross validation, we consider two model performance criteria: (1) predictive interval coverage (PIC) and  (2) rank probability score (RPS). PIC offers assessment of model adequacy, RPS enables model comparison.\\

\noindent{\bf Predictive Interval Coverage}\\

After the model is fitted to $\mathcal{S}^{train}$, the posterior predictive intensity function can supply posterior predictive point patterns and therefore samples from the posterior predictive distribution of $N(B_{k})$. For the $\ell$th posterior sample, $\ell = 1,...., L$, the associated predictive residual is defined as
\begin{align*}
R_{\ell}^{pred}(B_{k})=N^{test}(B_{k})-N^{(\ell)}(B_{k})
\end{align*}
where $N^{test}(B_{k})$ is the number of points of the test data in $B_{k}$.
If the model is adequate, the empirical predictive interval coverage rate, i.e., the proportion of intervals which contain $0$, is expected to be roughly the nominal level of coverage; below, we choose $90\%$ nominal coverage. Empirical coverage much less than the nominal suggests model inadequacy; predictive intervals are too optimistic.  Empirical coverage much above, for example $100\%$, is also undesirable.  It suggests that the model is introducing more uncertainty than needed.\\

\noindent{\bf Rank Probability Score}\\

\cite{GneitingRaftery(07)} propose the continuous rank probability score (CRPS).
This score is derived as a proper scoring rule and enables a criterion  for assessing the precision of a predictive distribution for continuous variables.
In our context, we seek to compare a predictive distribution to an observed count.
\cite{Czadoetal(09)} discuss rank probability scores (RPS) for count data.
Intuitively, a good model will provide a predictive distribution that is very concentrated around the observed count.
While the RPS has a challenging formal computational form, it is directly amenable to Monte Carlo integration.  In particular, for a given $B_{k}$, we calculate the RPS as
\begin{align*}
\text{RPS}(F, N^{test}(B_{k}))&=\frac{1}{L}\sum_{\ell=1}^{L}|N^{(\ell)}(B_{k})-N^{test}(B_{k})| \\
&-\frac{1}{2L^2}\sum_{\ell=1}^{L}\sum_{\ell^{'}=1}^{L}|N^{(\ell)}(B_{k})-N^{(\ell^{'})}(B_{k})|
\end{align*}
Summing over the collection of $B_{k}$ gives a model comparison criterion.  Smaller values of the sum are preferred.

\section{Data Analysis}

In this section, we implement our model first for simulated data and then for the SF crime dataset. We investigate our ability to recover the true model through the simulation study where we use the same geographic region and time period as in the real SF data.
We approximate the likelihood by taking space and circular time grid cells. We take 238 spatial grid cells for the full region and 64 spatial grid cells for the subregion.
As for the time grid, we take 48 time grid cells, i.e., 02:00--02:30, 02:30--03:00, $\ldots$, 25:30--02:00. As representative points to evaluate the intensity, we take the centroids of the cells. We consider a separable covariance specification for the full region and nonseparable covariance for the subregion around the Tenderloin and the northern part of the Mission districts.  We use the constructed spatial and temporal covariates as discussed in Section 4.
For the directional Gaussian kernels associated with the landmarks, we need to estimate the correlation parameters $\rho_{k}, k=1,2$.
We plug in the posterior means of these parameters obtained under the NHPP models for each crime category (see Table $\ref{tab:rho}$).

\begin{table}[htbp]
\caption{Posterior means of the correlation parameters $\rho_{1}$ and $\rho_{2}$ in the directional Gaussian kernels for the NHPP model.}
\begin{tabular}{lcccccccc}
\hline
  &  \multicolumn{3}{c}{Full region} & & \multicolumn{3}{c}{Subregion}   \\
\hline
  & Assault & Burg/Rob & Drug & & Assault & Burg/Rob & Drug  \\
\hline
$\rho_{1}$ & 0.097 & 0.088 &  0.209  && 0.151 & 0.232 &  0.284  \\
$\rho_{2}$ & -0.142 & -0.344 &  0.054 & & -0.537 & -0.893 &  0.946   \\
\hline
\end{tabular}
\label{tab:rho}
\end{table}

Following Section 4.1, the priors for the parameters are chosen as $\mu_{w}, \delta_{w} \sim \mathcal{G}(2, 0.05)$, $\bm{\beta}\sim \mathcal{N}(\bm{0}, 100\mathbf{I}_{K})$, $\sigma^2\sim \mathcal{IG}(2, 0.05)$, $\phi_{s}\sim \mathcal{U}[0, 0.3]$, $\phi_{t}\sim \mathcal{U}[0, 6]$ and $\gamma\sim \mathcal{U}[0,1)$.
We generate 150,000 posterior samples and take 100,000 as burn-in period for the full region and 100,000 samples and take 50,000 as burn-in period for the subregion.
We also report the inefficiency factor (IF)\footnote{The inefficiency factor is the ratio of the numerical variance of the estimate from the MCMC sample relative to that from hypothetical uncorrelated samples, and is defined as $1+2\sum_{s=1}^{\infty}\rho_{s}$ where $\rho_{s}$ is the sample autocorrelation at lag $s$. These values on tables are calculated with samples obtained at each 50 iteration} which suggests the relative number of correlated draws necessary to attain the same variance of the posterior sample mean from the uncorrelated draws (Chib (2001)). 
\subsection{Simulation study results}

The simulation data are generated from our model specification in section 4.1 with posterior means $\rho_{1}$ and $\rho_{2}$ for assault events (i.e., $\rho_{1}=0.097$ and $\rho_{2}=-0.142$) and without day of week effects, i.e., $\mu_{w}=\mu$ and $\delta_{w}=\delta$ for $w=Sun, \ldots, Sat$. Specifically, we generate a point pattern under a separable covariance function for the full region and a second point pattern under a nonseparable covariance function for the subregion.  The resulting numbers of points are 9,935 for the full region and 6,451 for the subregion.  A LGCP model with separable covariance (LGCP-Sep) and NHPP model with the spatial and time covariates are implemented for simulated data over the full region.
\begin{table}[htbp]
\caption{Estimation summary for NHPP and the space by circular time LGCP with separable covariance for the simulation data over the full region}
\begin{tabular}{lcccccccccc}
\hline
  &  \multicolumn{4}{c}{NHPP} & & \multicolumn{4}{c}{LGCP-Sep (True)}   \\
\hline
  & True & Mean &  $95\%$CI & IF & & True & Mean &  $95\%$CI & IF  \\
\hline
$\mu$ & 1 & 0.081 &  [0.079, 0.084] & 1 & $\mu$ & 1 & 1.317 &  [0.705, 2.485] & 72  \\
$\delta$ & 0.5 & 1.125 &  [1.049, 1.208] & 1 & $\delta$ & 0.5 & 0.391 &  [0.185, 0.649] & 66  \\
$\beta_{1}$ & 3 & 3.048 &  [2.966, 3.128] & 4  & $\beta_{1}$ & 3 & 3.990 &  [3.165, 4.949] & 72 \\
$\beta_{2}$ & 3 & 3.287 &  [3.204, 3.362] & 4  & $\beta_{2}$ & 3 & 2.673 &  [1.599, 3.907] & 74  \\
$\rho_{1}$ & 0.097 & 0.551 &  [0.501, 0.596] & 1 & $\sigma^2$ & 3 & 3.256 &  [2.850, 3.886] & 73 \\
$\rho_{2}$ & -0.142 & -0.424 &  [-0.480, -0.362] & 1  &  $\phi_{s}$ & 0.02 & 0.019 &  [0.018, 0.021] & 60 \\
 &  &  &   &   & $\phi_{t}$ & 0.1 & 0.101 &  [0.085, 0.117] & 65 \\
 &  &  &   &   & $\sigma^2\phi_{s}$ & 0.06 & 0.064 &  [0.056, 0.075] & 68 \\
 &  &  &   &   & $\sigma^2\phi_{t}$ & 0.3 & 0.328 &  [0.290, 0.375] & 56 \\
\hline
\end{tabular}
\label{tab:result_Sim_F}
\end{table}

Table $\ref{tab:result_Sim_F}$ shows the estimation summary for the simulation data over the full region. The NHPP inference shows high precision but poor accuracy.
For LGCP-Sep, although consistency for $\phi_{s}$ and $\sigma^2$ is not guaranteed (see \cite{Zhang(04)}), we see good recovery of the parameters.
For the NHPP model with the spatial and time covariates, since the space time Gaussian processes are not included in the model, the estimation results have biases relative to the true values based on the LGCP-Sep.
The true and posterior intensity surface at two time grids: (1) 12:00--12:30 and (2) 24:00--24:30 are shown in the supplementary material. Compared with the true surface, we see some preference for the LGCP-Sep.

\begin{table}[htbp]
\caption{Estimation summary for the space by circular time LGCP for simulation data over the subregion: LGCP with separable covariance (left) and LGCP with nonseparable covariance (right)}
\begin{tabular}{lcccccccccc}
\hline
  &  \multicolumn{4}{c}{LGCP-Sep} & & \multicolumn{4}{c}{LGCP-NonSep (True)}   \\
  & True & Mean &  $95\%$CI & IF & & True & Mean &  $95\%$CI & IF  \\
\hline
$\mu$ & 1 & 0.778 &  [0.196, 2.283] & 73 & $\mu$ & 1 & 0.533 &  [0.144, 1.796] & 73 \\
$\delta$ & 0.5 & 0.196 &  [0.000, 0.502] & 70 & $\delta$ & 0.5 & 0.249 &  [0.076, 0.491] & 58  \\
$\beta_{1}$ & 3 & 2.909 &  [1.913, 4.150] & 73 & $\beta_{1}$ & 3 & 2.961 &  [2.261, 3.946] & 72  \\
$\beta_{2}$ & 3 & 2.600 &  [1.903, 3.213] & 71 & $\beta_{2}$ & 3 & 2.472 &  [1.895, 3.132] & 70  \\
$\sigma^2$ & 1 & 0.842 &  [0.567, 1.442] & 72 & $\sigma^2$ & 1 & 0.913 &  [0.535, 1.301] & 72 \\
$\phi_{s}$ & 0.02 & 0.029 &  [0.014, 0.043] & 70 & $\phi_{s}$ & 0.02 & 0.024 &  [0.014, 0.039] & 69  \\
$\phi_{t}$ & 0.1 & 0.168 &  [0.117, 0.236] & 58 &  $\phi_{t}$ & 0.1 & 0.125 &  [0.084, 0.181] & 61  \\
$\sigma^2\phi_{s}$ & 0.02 & 0.023 &  [0.017, 0.032] & 64 & $\sigma^2\phi_{s}$ & 0.02 & 0.022 &  [0.013, 0.032] & 67  \\
$\sigma^2\phi_{t}$ & 0.1 & 0.139 &  [0.091, 0.217] & 64 & $\sigma^2\phi_{t}$ & 0.1 & 0.111 &  [0.075, 0.163] & 61  \\
 &  &  &  &  & $\gamma$ & 0.8 & 0.498 &  [0.031, 0.951] & 67  \\
\hline
\end{tabular}
\label{tab:result_Sim_S}
\end{table}

Table $\ref{tab:result_Sim_S}$ shows the estimation summary for the simulated data over the subregion.
True values of parameters are recovered well by the LGCP model with nonseparable covariance (LGCP-NonSep). Although the true value of $\gamma$ is included in the $95\%$ CI, the posterior for $\gamma$ has large variance.
Since the true model is LGCP-NonSep with $\gamma=0.8$, the estimation of $\phi_{t}$ by LGCP-Sep yields some bias.
The true and posterior intensity surface for two time grids: (1) 12:00--12:30 and (2) 24:00--24:30 are shown in supplemental material.
The estimated intensity surfaces for LGCP-Sep and LGCP-NonSep are similar to each other. Additionally, we also compare the values of RPS and PIC of LGCP-Sep with those of LGCP-NonSep. Model comparison will change with the selection of time grid but, altogether, the differences are ignorable. Details are provided in the supplementary material.

\subsection{Real Data Application}
\leavevmode \\
\noindent{\bf Full region}\\

The numbers of crime events in 2012 for the full region are 9834 for assault, 9884 for burglary/robbery and 6234 for drug. We implement LGCP-Sep and NHPP, assessing validation with $90\%$ PIC and comparison using RPS. We separate $\mathcal{S}^{test}$ and $\mathcal{S}^{train}$ with $p=0.5$.
We calculate RPS and $90\%$ PIC for three time ranges: (1) 02:00--10:00, (2) 10:00--18:00 and (3) 18:00--02:00.  As for the choice of $B_{k}$, from the grid approximation over space and time, each $B_{k}$ is chosen as a sum of grid units over space for each time interval. As above, the area of $B_{k}$ is approximately equal to $q|D|$ where $|D|$ is the total area and here we choose the relative size $q \leq 0.1$. We randomly choose 1,000 sets of $B_{k}$ uniformly over $D$ and, following Section 4.2, compute average RPS and $90\%$ PIC over these sets.
Figure \ref{fig:MV_F} shows RPS and PIC for the three crime categories and three time ranges.
Figure \ref{fig:MV_F} reveals that the LGCP-Sep model outperforms the NHPP model for all of the crime types and all of the time ranges. Figure \ref{fig:MV_F} also demonstrates that the LGCP-Sep $90\%$ predictive intervals capture nominal coverage very well while the PIC's for the NHPP are too small.

There may be interest in assessing \emph{local} model adequacy to see if there are areas where the model is not performing well.  Due to the flexibility of the LGCP-Sep model, we would not expect to see any local anomalies.  However, we can assess this using our posterior samples and our collection of $B_{k}$'s, enabing calculation of local PIC's.  These are presented in Figure $\ref{fig:PIC_F}$ for each crime type over the spatial region. We see that most are above the nominal, a few below, but there is no evident pattern, no clustering.  So, we can conclude that any over- or under-fitting occurs at random over the region.

\begin{figure}[htbp]
  \caption{The rank probability score (left) and predictive interval coverage (right) for the full region: NHPP (blue line) and LGCP with separable covariance (red line). $q$ is the relative area of $B_{k}$ to that of $D$.}
 \begin{minipage}{0.48\hsize}
  \begin{center}
   \includegraphics[width=6.5cm]{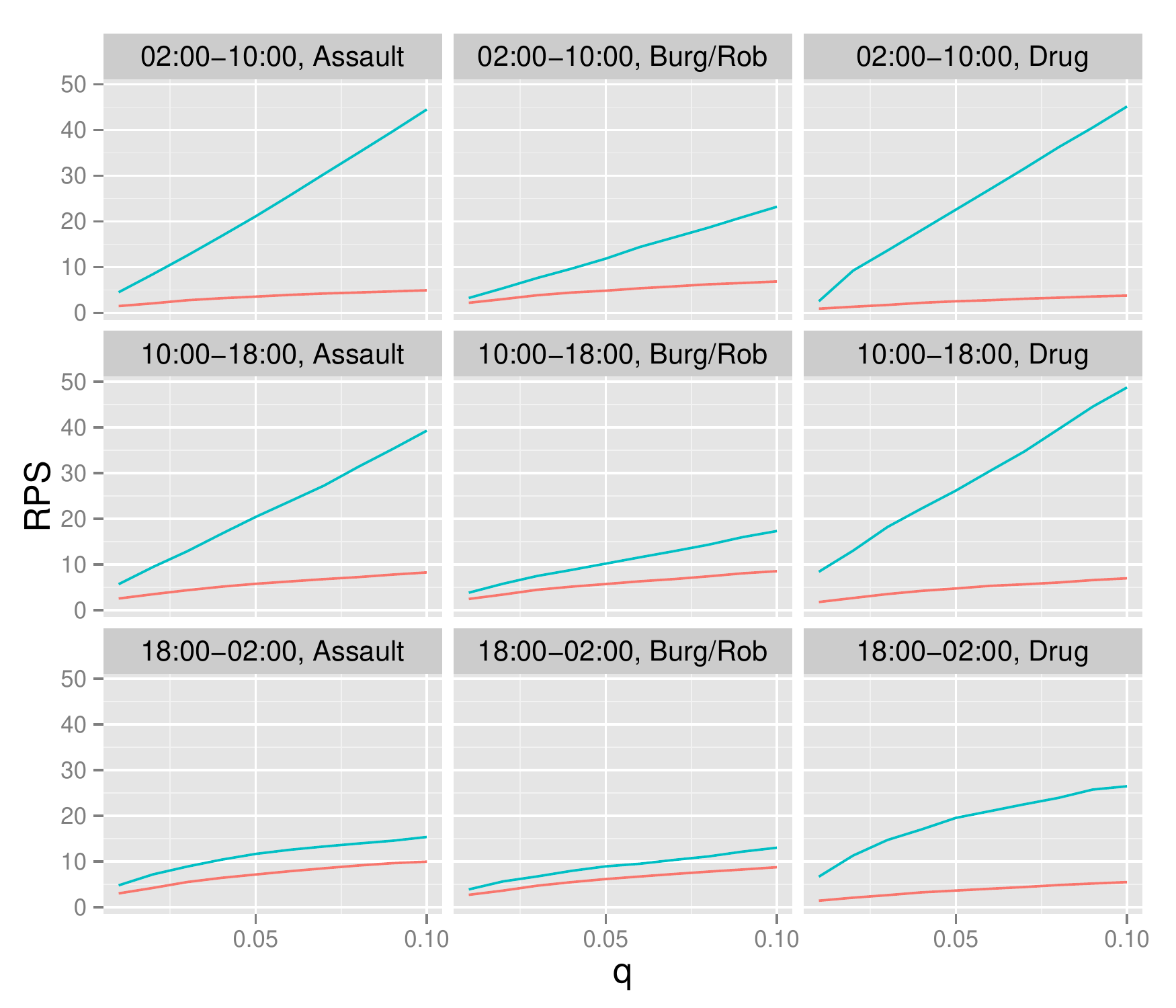}
  \end{center}
\end{minipage}
 \hfill
 \hfill
 \begin{minipage}{0.48\hsize}
  \begin{center}
   \includegraphics[width=6.5cm]{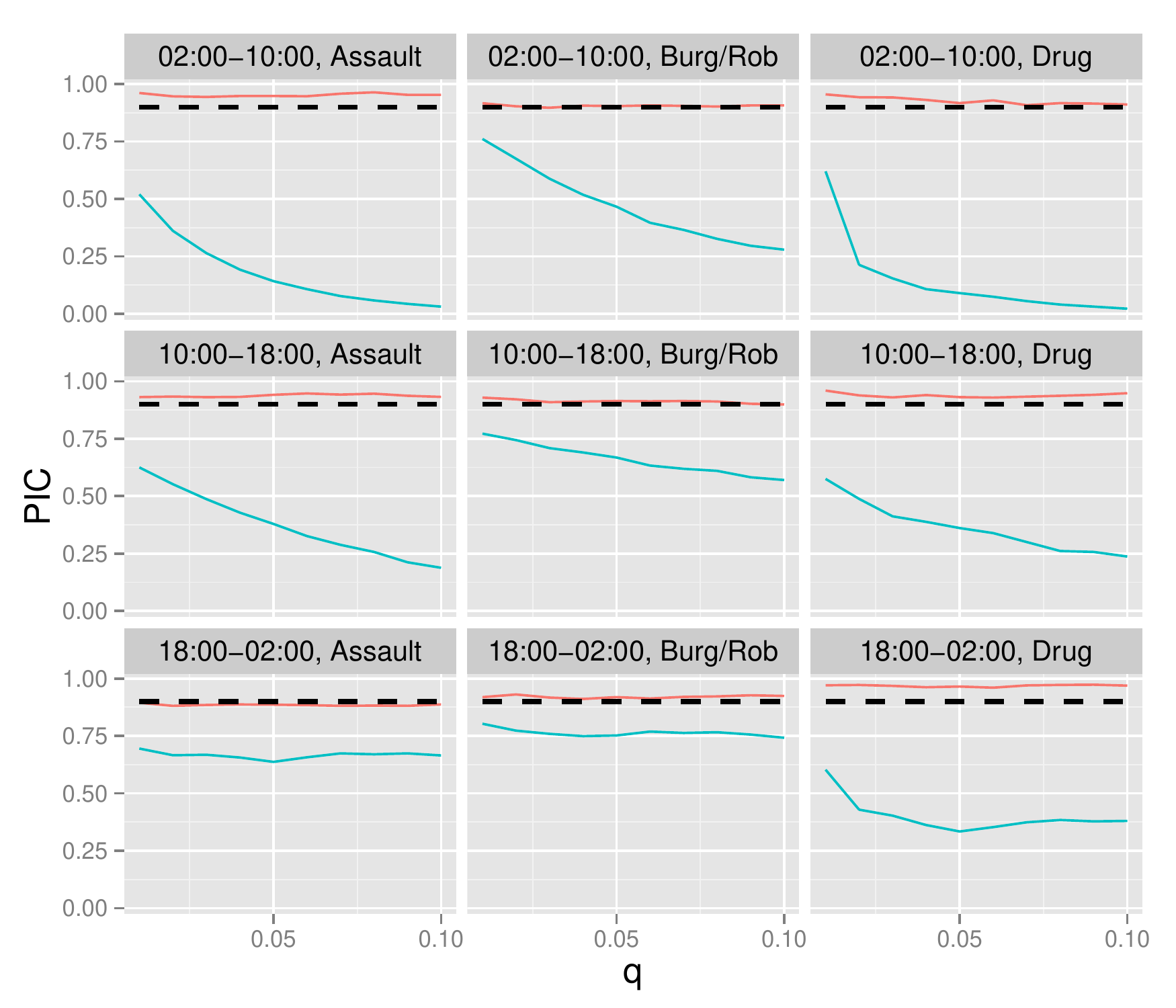}
  \end{center}
\end{minipage}
  \label{fig:MV_F}
\end{figure}

\begin{figure}[htbp]
  \caption{Local in-sample PIC for the three crime types}
  \begin{center}
   \includegraphics[width=13cm]{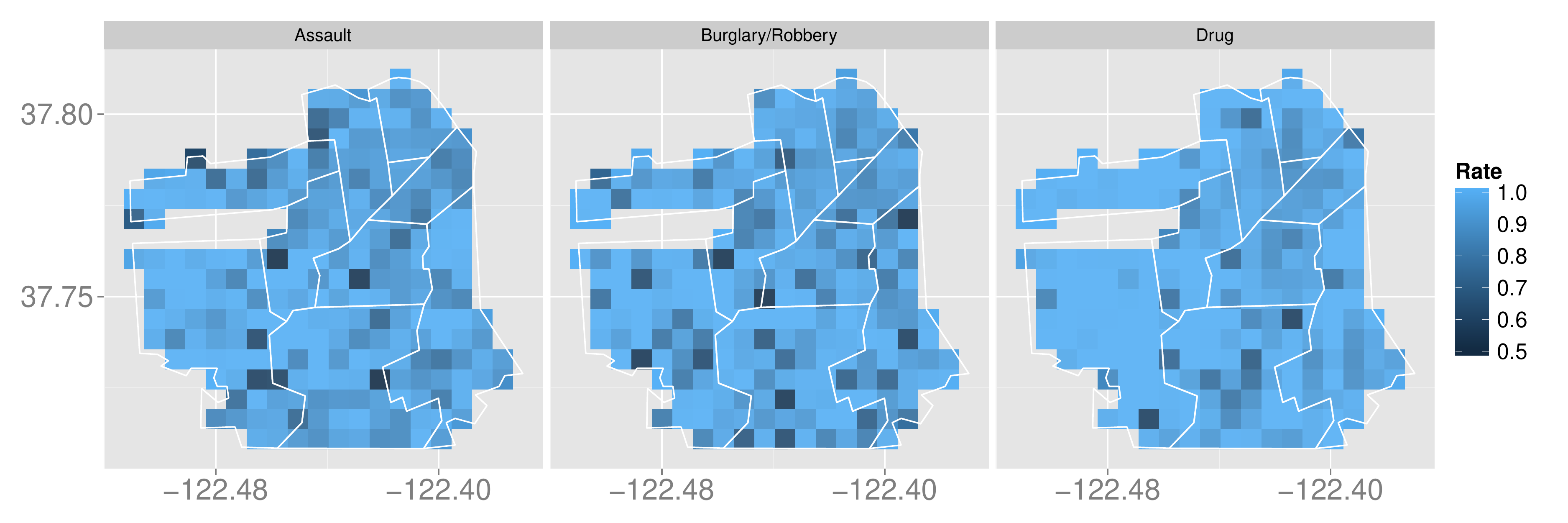}
  \end{center}
  \label{fig:PIC_F}
\end{figure}

Table \ref{tab:result_F} shows the estimation results for the space by circular time LGCP for the three crime categories in 2012. With day of week specific $\mu_{w}$ and $\delta_{w}$, $\mu_{0}$ and $\delta_{0}$ are set to the means of them over the days of week, yielding $\mu_{w}-\mu_{0}$ and $\delta_{w}-\delta_{0}$ as deviations.
See Figure \ref{fig:Scale_F} below for inference on the $\delta_{w}$ across day of the week.
The spatial covariates $\bm{\beta}$ are \emph{positively} significant. In particular, $\beta_{1}$ and $\beta_{2}$ for drug crimes show larger values than those for the other crime types. This result suggests that drug events are more concentrated around landmarks $L_{1}$ and $L_{2}$.

\begin{table}[htbp]
\caption{Estimation results for space by circular time LGCP for the full region with separable covariance: assault (left), burglary/robbery (middle) and drug (right)}
\scalebox{0.9}[0.9]{
\begin{tabular}{lccccccccccccc}
\hline
  &  \multicolumn{3}{c}{Assault} & \multicolumn{3}{c}{Burglary/Robbery} & \multicolumn{3}{c}{Drug}  \\
  & Mean &  $95\%$CI & IF & Mean &  $95\%$CI & IF & Mean &  $95\%$CI & IF \\
\hline
$\mu_{0}$ & 36.94 &  [20.62, 60.12] & 70 & 36.18 &  [21.45, 57.79] & 70 & 32.43 &  [19.04, 58.32] & 69 \\
$\delta_{0}$ & 0.342 &  [0.201, 0.613] & 68 & 0.188 &  [0.069, 0.424] & 69 & 0.344 &  [0.137, 0.672] & 70 \\
$\beta_{1}$ & 1.654 &  [1.146, 2.023] & 69 & 2.646 &  [2.038, 3.542] & 73 & 3.874 &  [2.146, 5.045] & 74 \\
$\beta_{2}$ & 1.202 &  [0.614, 1.778] & 70 & 0.470 &  [0.048, 0.823] & 65 & 3.745 &  [2.117, 4.399] & 71 \\
$\sigma^2$ & 5.598 &  [5.064, 6.471] & 73 & 5.756 &  [5.331, 6.219] & 72 & 8.424 &  [7.868, 8.984] & 67 \\
$\phi_{s}$ & 0.011 &  [0.010, 0.013] & 70 & 0.005 &  [0.005, 0.007] & 70 & 0.027 & [0.025, 0.030] & 68 \\
$\phi_{t}$ & 0.137 &  [0.123, 0.151] & 58 & 0.178 &  [0.163, 0.196] & 59 & 0.161 & [0.140, 0.182] & 63 \\
$\sigma^2\phi_{s}$ & 0.066 &  [0.059, 0.072] & 61 & 0.033 &  [0.028, 0.037] & 65 & 0.231 & [0.207, 0.259] & 63 \\
$\sigma^2\phi_{t}$ & 0.769 &  [0.681, 0.864] & 61 & 1.027 &  [0.887, 1.164] & 65 & 1.362 & [1.182, 1.540] & 60 \\
\hline
\end{tabular}
}
\label{tab:result_F}
\end{table}

Figure \ref{fig:Scale_F} shows the posterior mean and 95$\%$ CI of $\sum_{j=1}^{J} \lambda(\bm{s}_{j}^{*},t_{j}^{*}, w)\Delta_{s,t,w}$ against counts on each day of week.
For a given $w$, $\sum_{j=1}^{J} \lambda(\bm{s}_{j}^{*},t_{j}^{*}, w_{j}^{*})\Delta_{s,t,w}$ is approximately the expected number of crime events on day $w$ a year.
The left panel demonstrates that the posterior mean of $\sum_{j=1}^{J} \lambda(\bm{s}_{j}^{*},t_{j}^{*}, w)\Delta_{s,t,w}$ traces the observed counts on days of week accurately.
The right panel displays the posterior mean and 95$\%$ CI of $\delta_{w}$. Although the variance of $\delta_{w}$ is large, this figure shows that $\delta_{w}$ varies with day of week; for assault, weekend $\delta$'s are larger. Since all of the $\delta_{w}$'s are positive, regardless of day of week or type of crime, we find elevated risk in the evening hours. Interestingly, although drug counts on Wednesday are larger than those on other days, $\delta_{Wed}$ for drug is smaller than those for the other days.
Additionally, in the supplemental material, we provide figures for the posterior mean intensity surfaces for the three crime categories for three time grids: (1) 08:00--08:30, (2) 16:00--16:30 and (3) 24:00--24:30.
The figures reveal different intensity patterns for each category and time grid.

\begin{figure}[htbp]
  \caption{Posterior mean and 95$\%$ CI of $\sum_{j=1}^{J} \lambda(\bm{s}_{j}^{*},t_{j}^{*}, w)\Delta_{s,t,w}$ (left: dotted points are crime counts) and $\delta_{w}$ (right) on each day of week for the full region: dashed lines are 95$\%$ CI}
 \begin{minipage}{0.48\hsize}
  \begin{center}
   \includegraphics[width=6.5cm]{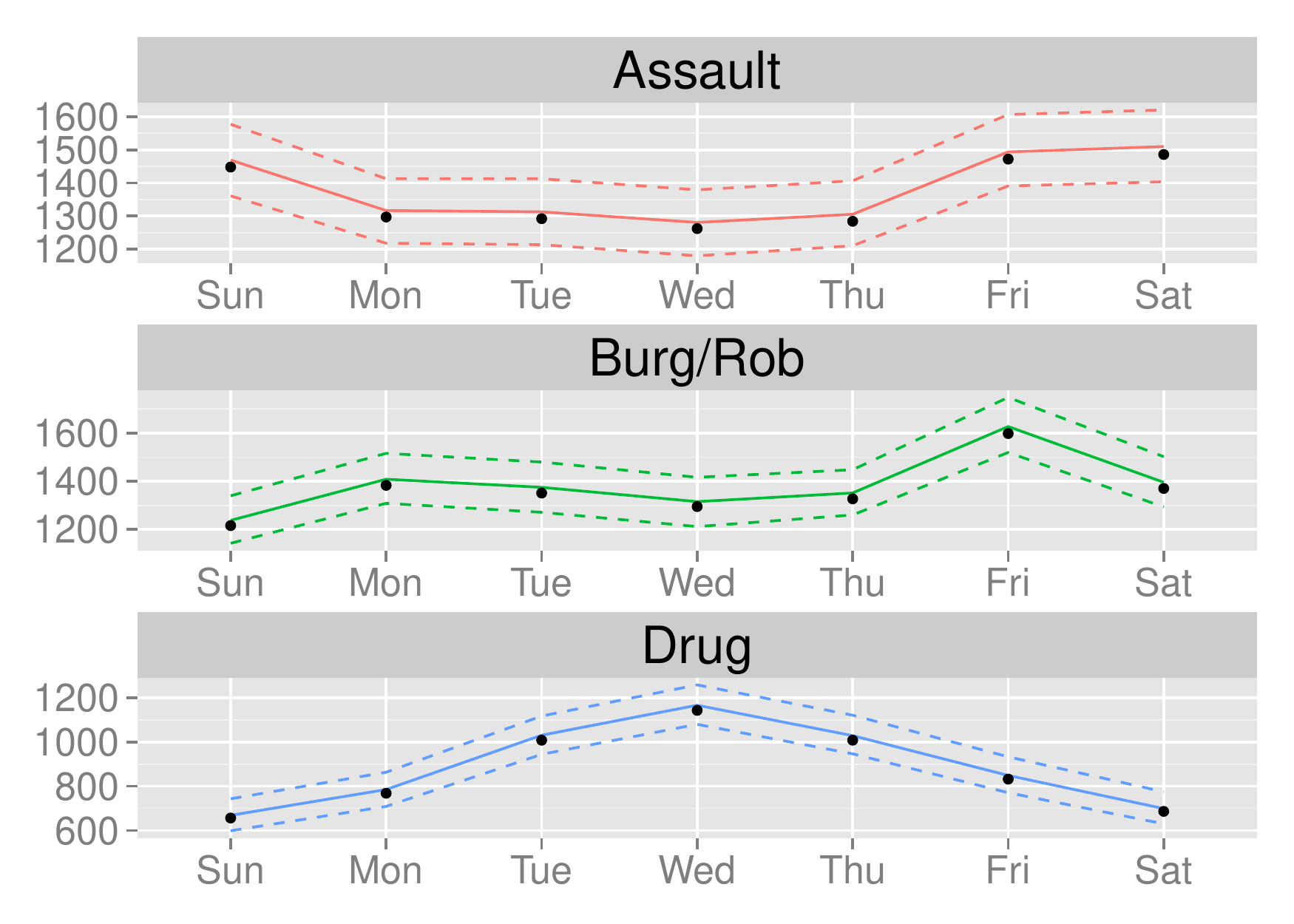}
  \end{center}
 \end{minipage}
 \hfill
 \begin{minipage}{0.48\hsize}
  \begin{center}
   \includegraphics[width=6.5cm]{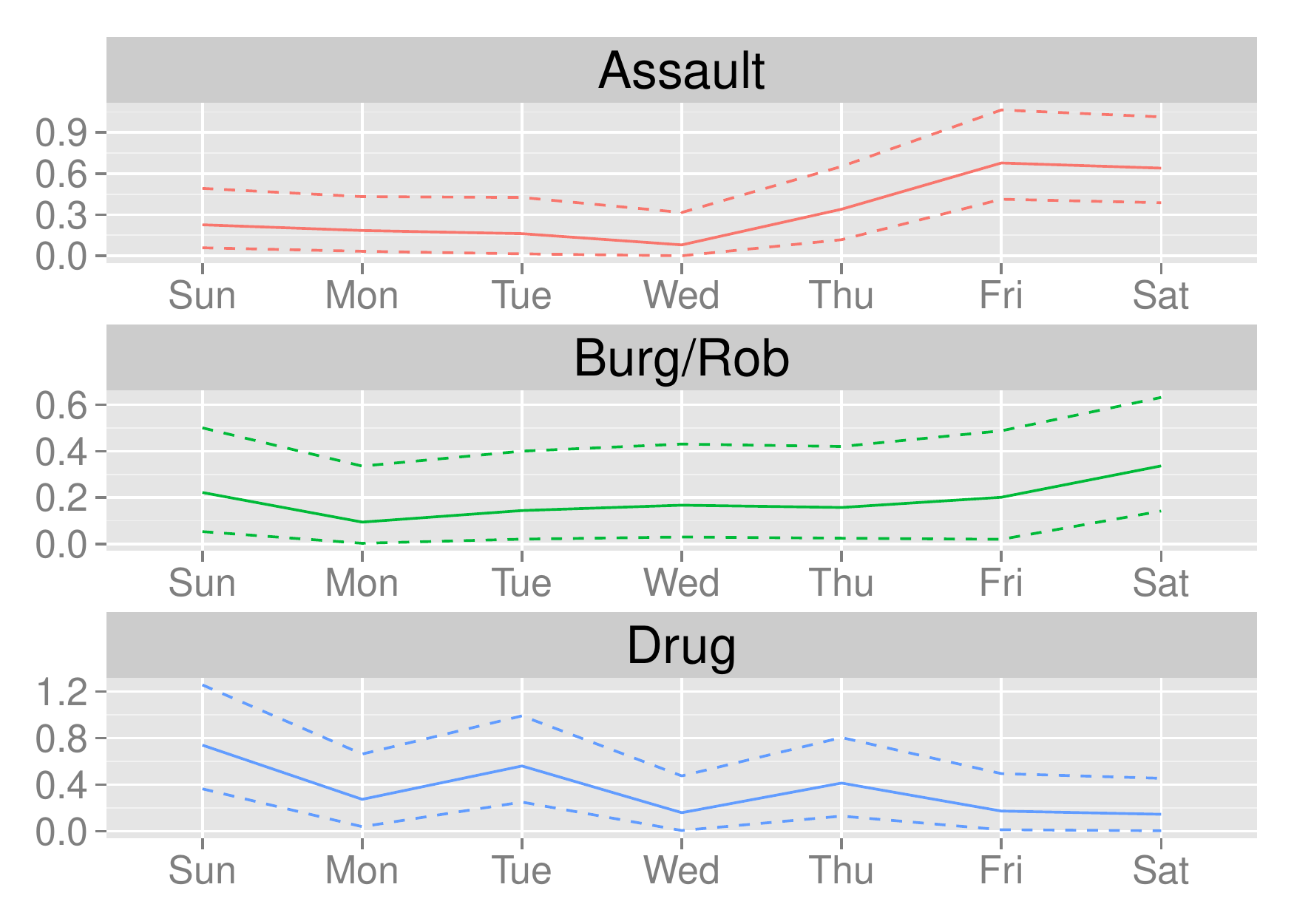}
  \end{center}
 \end{minipage}
  \label{fig:Scale_F}
\end{figure}

\noindent{\bf Subregion}\\

Finally, we turn to the nonseparable case, providing results for the subregion and comparison with the separable case.
The number of points are 5579 for assault, 5407 for burglary/robbery and 4415 for drug crimes.
Figure \ref{fig:MV_S} shows the RPS and PIC for three models for the subregion: (1) NHPP, (2) LGCP-Sep and (3) LGCP-NonSep. Although both LGCP models fit considerably better than NHPP model, the model performance of LGCP-Sep is difficult to be distinguish from that of LGCP-NonSep. This result is consistent with our findings in the simulation study.

\begin{figure}[htbp]
  \caption{The rank probability score (left) and predictive interval coverage (right) for the subregion: NHPP (blue line), LGCP with separable covariance (red line) and LGCP with nonseparable covariance (green line). $q$ is the relative area of $B_{k}$ to that of $D$.}
 \begin{minipage}{0.48\hsize}
  \begin{center}
   \includegraphics[width=6.5cm]{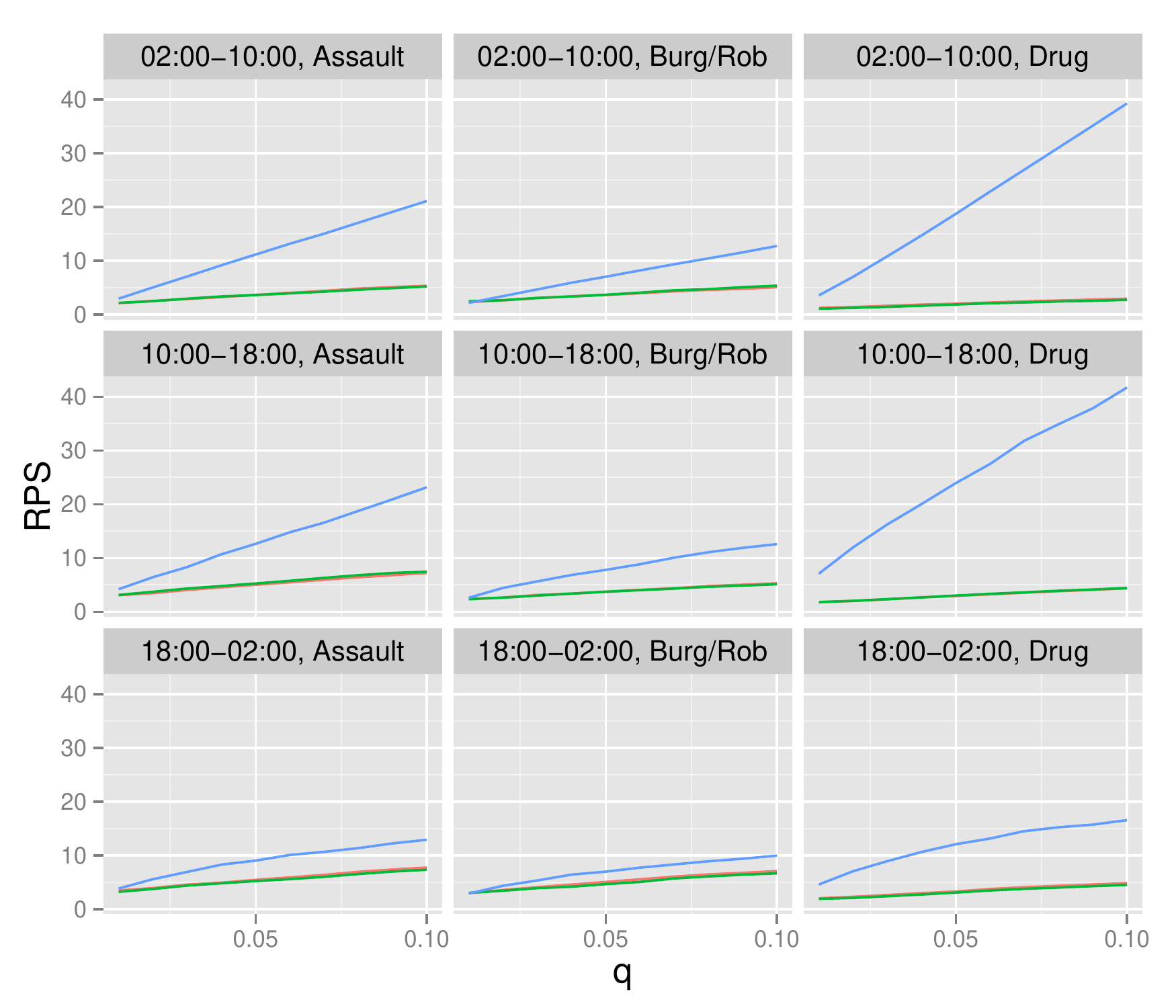}
  \end{center}
\end{minipage}
 \hfill
 \hfill
 \begin{minipage}{0.48\hsize}
  \begin{center}
   \includegraphics[width=6.5cm]{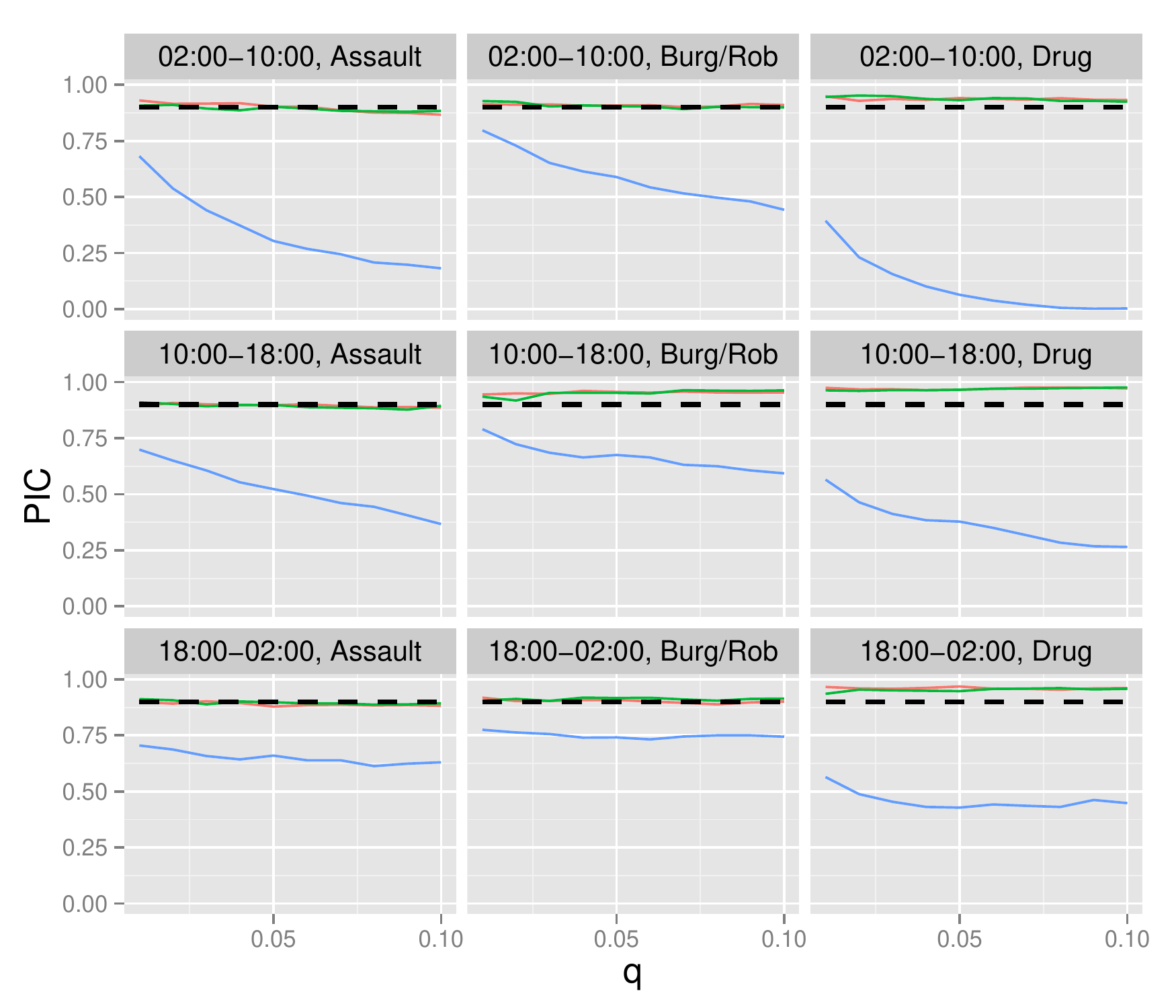}
  \end{center}
\end{minipage}
  \label{fig:MV_S}
\end{figure}

Table \ref{tab:result_S} shows the estimation results.
The estimated values of $\mu_{0}$ for assault and burglary/robbery crimes are higher and have larger variances than that for drug crimes.
From our model specification, $\mu_{0}$ and $\sigma^2$ have strong positive correlation because both parameters are scale parameters for the intensity, i.e., $\mu_{0}\exp(-\sigma^2/2)$.
In fact, the results reveal larger values of $\mu_{0}$ and $\sigma^2$ for assault and burglary/robbery events, a smaller value of $\mu_{0}$ and $\sigma^2$ for drug events. $\gamma>0$ express the degree of nonseparability. It varies with crime type but has high uncertainty so that the differences between types are not distinguished. As in the simulation study, the difference between LGCP-Sep and LGCP-NonSep is very small with respect to RPS.
Figure \ref{fig:Scale_S} shows the posterior mean and 95$\%$ CI of the $\sum_{j=1}^{J} \lambda(\bm{s}_{j}^{*},t_{j}^{*}, w)\Delta_{s,t,w}$ against counts on each day of week.
The right figure exhibits the posterior mean and 95$\%$ CI of the $\delta_{w}$.
Although the results are similar to those for the full region, $\delta_{Wed}$ for drug is larger than that for the full region while now, $\delta_{Tue}$ for drug is smaller than that for the full region.

\begin{table}[htbp]
\caption{Estimation results for the space by circular time LGCP for the subregion with nonseparable covariance: assault (left), burglary/robbery (middle) and drug (right)}
\scalebox{0.9}[0.9]{
\begin{tabular}{lccccccccccccc}
\hline
  &  \multicolumn{3}{c}{Assault} & \multicolumn{3}{c}{Burglary/Robbery} & \multicolumn{3}{c}{Drug}  \\
  & Mean &  $95\%$CI & IF & Mean &  $95\%$CI & IF & Mean &  $95\%$CI & IF \\
\hline
$\mu_{0}$ & 38.49 &  [19.89, 66.82] & 67 & 67.93 &  [12.39, 51.44] & 71 & 0.281 &  [0.237, 0.764] & 71 \\
$\delta_{0}$ & 0.360 &  [0.196, 0.625] & 67 & 0.347 &  [0.000, 0.980] & 71 & 0.284 &  [0.001, 0.764] & 70 \\
$\beta_{1}$ & 1.875 &  [0.819, 2.586] & 72 & 2.236 &  [1.819, 2.551] & 65 & 4.204 &  [3.643, 4.975] & 71 \\
$\beta_{2}$ & 1.889 &  [1.103, 2.726] & 72 & 1.498 &  [0.449, 2.464] & 66 & 2.562 &  [1.451, 3.972] & 66 \\
$\sigma^2$ & 5.065 &  [4.435, 5.508] & 70 & 7.244 &  [6.661, 8.382] & 72 & 2.331 &  [1.761, 2.935] & 67 \\
$\phi_{s}$ & 0.014 &  [0.012, 0.016] & 65 & 0.003 &  [0.003, 0.004] & 60 & 0.075 & [0.059, 0.094] & 66 \\
$\phi_{t}$ & 0.147 &  [0.117, 0.191] & 69 & 0.236 &  [0.196, 0.277] & 59 & 0.604 & [0.354, 0.815] & 68 \\
$\gamma$ & 0.104 &  [0.006, 0.246] & 55 & 0.069 &  [0.003, 0.194] & 48 & 0.236 & [0.016, 0.533] & 53 \\
$\sigma^2\phi_{s}$ & 0.072 &  [0.063, 0.083] & 57 & 0.033 &  [0.028, 0.037] & 65 & 0.231 & [0.207, 0.259] & 63 \\
$\sigma^2\phi_{t}$ & 0.742 &  [0.617, 0.933] & 66 & 1.027 &  [0.887, 1.164] & 65 & 1.362 & [1.182, 1.540] & 60 \\
\hline
\end{tabular}
}
\label{tab:result_S}
\end{table}

\begin{figure}[htbp]
  \caption{Posterior mean and 95$\%$ CI of $\sum_{j=1}^{J} \lambda(\bm{s}_{j}^{*},t_{j}^{*}, w)\Delta_{s,t,w}$ (left: dotted points are crime counts) and $\delta_{w}$ (right) on each day of week for the subregion: dashed lines are 95$\%$ CI}
 \begin{minipage}{0.48\hsize}
  \begin{center}
   \includegraphics[width=6.5cm]{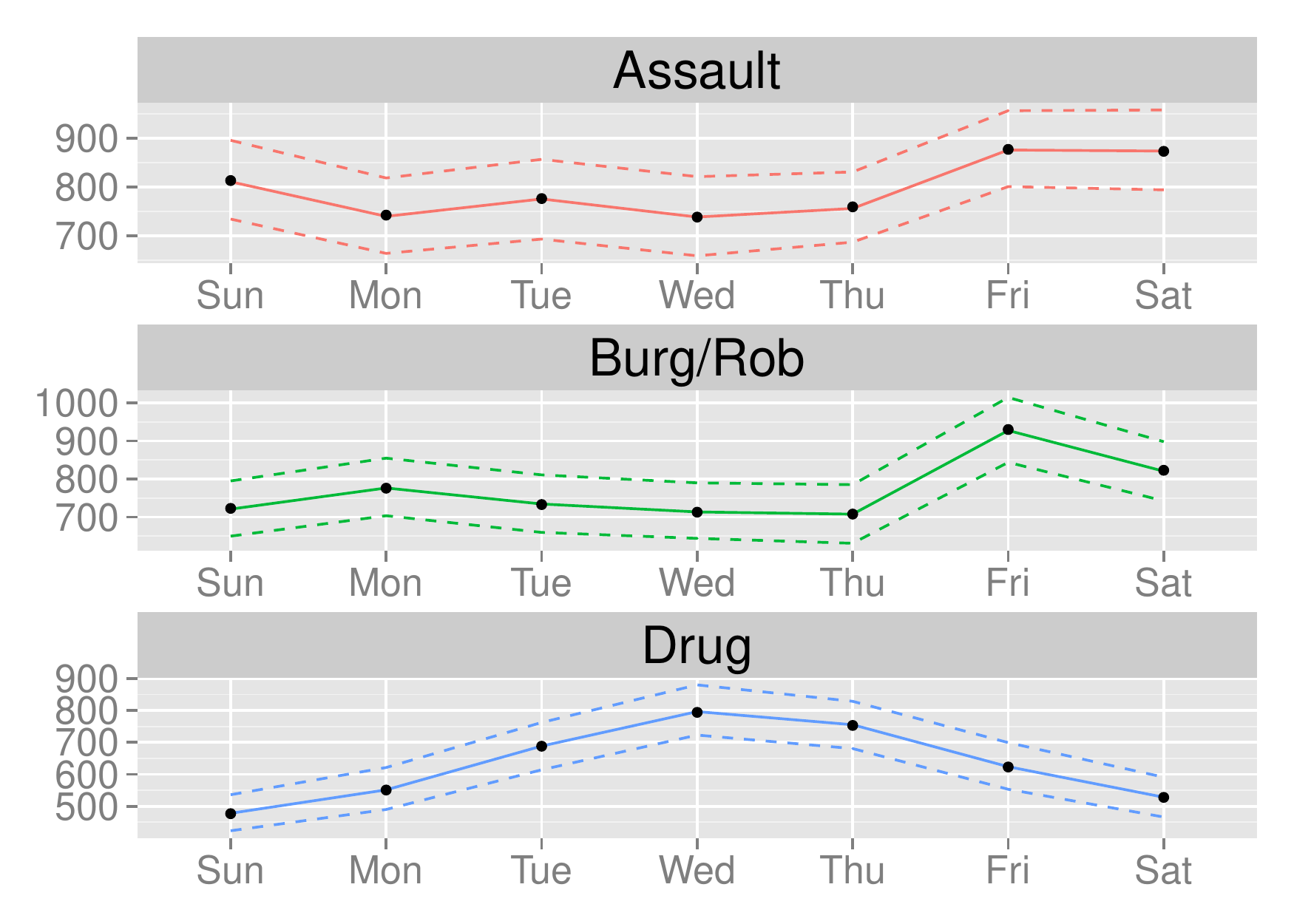}
  \end{center}
 \end{minipage}
 \hfill
 \begin{minipage}{0.48\hsize}
  \begin{center}
   \includegraphics[width=6.5cm]{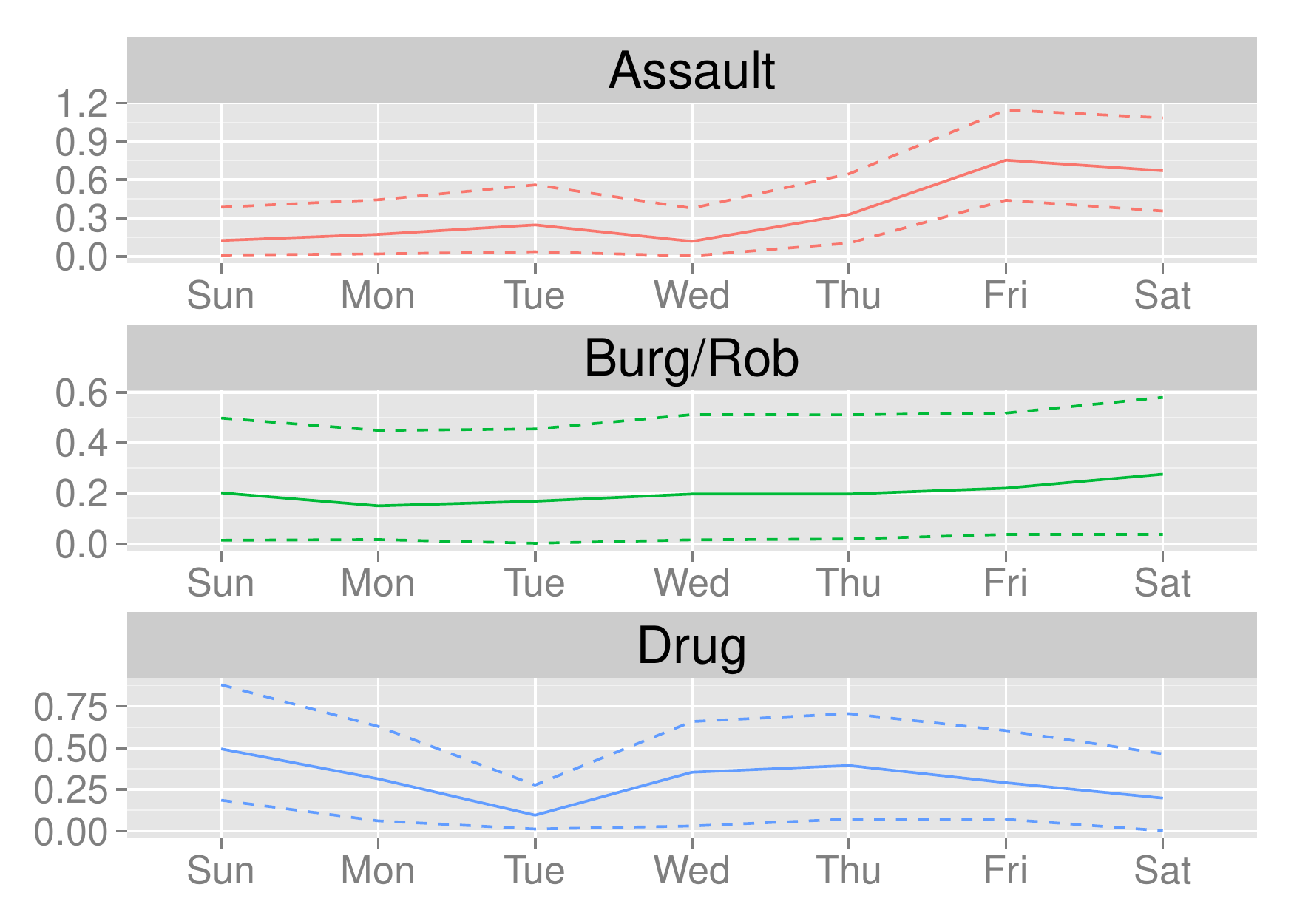}
  \end{center}
 \end{minipage}
  \label{fig:Scale_S}
\end{figure}

\section{Summary and future work}
We have looked at times and locations of crime events for the city of San Francisco.  We have argued that these data should be treated as point patterns in space and time where time should be treated as circular.  We introduced derived spatial covariates (using distance from landmarks) and temporal covariates (using day of the week). We have looked at NHPP and LGCP models for such data. For the latter, we have proposed valid space and circular time Gaussian processes, both separable and nonseparable, for use in the  LGCP.  We have shown through a simulation example, that we can recover the underlying model and intensity surface. We have discussed criteria for model adequacy (PIC) and model comparison (RPS).  We have shown that the LGCP outperforms the NHPP for the SF crime data. However, strong support for nonseparability for the subregion is not seen through our model estimation.

Future work will focus on more efficient computation.  It  will find us trying to develop appropriate approximations to enable us to fit the nonseparable model to larger regions.
It will also consider alternative approaches to the likelihood approximation, following strategies proposed by \cite*{AdamsMurrayMacKay(09)}.

\section*{Acknowledgements}

The work of the first author was supported in part by the Nakajima Foundation.
The authors thank Giovanna Jona Lasinio for suggesting this problem, for useful conversations, and for providing the San Francisco dataset.
The computational results are obtained by using Ox (\cite{Doornik(06)}).

\begin{supplement}
\sname{Supplement to "Space and circular time log Gaussian Cox processes with application to crime event data"}\label{suppA}
\slink[url]{url}
\sdescription{In this online supplement article, we provide (1) proof of the validity of our proposed nonseparable covariance function on $\mathbb{R}^2\times S^1$ and (2) additional figures and tables to see posterior mean intensity estimates under different models.}
\end{supplement}

\newpage
\bibliographystyle{imsart-nameyear}
\bibliography{SP}

\end{document}